\documentclass[10pt,journal,compsoc, review]{IEEEtran}

\usepackage{mathptmx}
\usepackage{graphicx}
\graphicspath{ {./EduVis_Figures/} }

\usepackage{caption}
\usepackage{subcaption}
\usepackage{times}
\usepackage{todonotes}
\usepackage{xcolor}
\usepackage{color}
\usepackage{CJKutf8}
\usepackage[fixed]{fontawesome5}
\usepackage{enumerate}
\usepackage{ulem}
\usepackage{url}
\usepackage{tabu}                      
\usepackage{booktabs}                  
\usepackage{kotex}
\usepackage{comment}

\ifCLASSOPTIONcompsoc
  \usepackage[nocompress]{cite}
\else
  \usepackage{cite}
\fi

\newcommand{\cleanRevision}[1]
{\textcolor{blue}{\textbf{#1}}}

\definecolor{orange}{rgb}{1,0.5,0}

\definecolor{capri}{rgb}{0.0, 0.75, 1.0}

\definecolor{cocoabrown}{rgb}{0.82, 0.41, 0.12}

\definecolor{darkcyan}{rgb}{0.0, 0.55, 0.55}

\definecolor{darkmagenta}{rgb}{0.55, 0.0, 0.55}

\ifCLASSINFOpdf
 
\else
 
\fi

\begin{document}
%
\title{HisVA: A Visual Analytics System \\ for Studying History  }

\author{
    Dongyun Han$^1$, \thanks{$^1$ This research was mainly performed when Dongyun Han was with UNIST}
    Gorakh Parsad,
    Hwiyeon Kim,
    Jaekyom Shim,
    Oh-Sang Kwon,
    Kyung A Son,\\
    Jooyoung Lee,
    Isaac Cho,
    and Sungahn Ko$^2$\thanks{$^2$ Corresponding author}

\IEEEcompsocitemizethanks{\IEEEcompsocthanksitem 
Dongyun Han and Isaac Cho are with Utah State University. E-mail:{{dongyun.han}@aggiemail.usu.edu, {isaac.cho}@usu.edu} \protect\\

\IEEEcompsocthanksitem  Gorakh Parsad, Hwiyeon Kim, Jaekyom Shim, Oh-Sang Kwon, Kyung A Son, Jooyoung Lee, and Sungahn Ko are with UNIST. E-mail:{gnldus28, badyalgaurav, jaekyom, oskwon, kasohn, shallibrown, sako}@unist.ac.kr.}
\thanks{Manuscript received April 19, 2005; revised August 26, 2015.}    
}

\IEEEtitleabstractindextext{
\begin{abstract}
Studying history involves many difficult tasks. Examples include searching for proper data in a large event space, understanding stories of historical events by time and space, and finding relationships among events that may not be apparent. Instructors who extensively use well-organized and well-argued materials (e.g., textbooks and online resources) can lead students to a narrow perspective in understanding history and prevent spontaneous investigation of historical events, with the students asking their own questions. In this work, we proposed HisVA, a visual analytics system that allows the efficient exploration of historical events from Wikipedia using three views: event, map, and resource. 
HisVA provides an effective event exploration space, where users can investigate relationships among historical events by reviewing and linking them in terms of space and time. 
To evaluate our system, we present two usage scenarios, a user study with a qualitative analysis of user exploration strategies, and 
in-class deployment results.
\end{abstract}

\begin{IEEEkeywords}
Visualization for Education, Event Visualization, Studying History, Wikipedia
\end{IEEEkeywords}
}

\maketitle

\IEEEdisplaynontitleabstractindextext

\IEEEpeerreviewmaketitle

\IEEEraisesectionheading{
\section{Introduction}
\label{sec:introduction}}

\IEEEPARstart{I}{n} a conventional history class, instructors deliver lectures based on textbooks, which contain well-organized mainstream arguments in the academic field of history. Such lecture-oriented history classes are still preferred, due to their advantages in conveying historical knowledge to students ~\cite{levesque2015history}; however, the strength of the conventional class implies that students acquire knowledge passively from textbooks and instructors, and seldom investigate and analyze historical events and contexts by themselves~\cite{mccarthy2000active}. As a result, students in conventional classes are less likely to study how to construct knowledge on their own and to develop their own perspectives on historical events~\cite{brooks1999search}.

A growing number of history instructors have begun to emphasize the importance of developing students’ critical thinking ability and their own historical perspectives during the historical knowledge acquisition process~\cite{clark2017surprise, Seefeldt09}. This trend is in line with constructivism education theories~\cite{richardson2003constructivist,lebow1993constructivist, driscoll1994psychology}, which emphasize that students should be given opportunities to ``construct'' their own meaning of knowledge according to their interests and experiences. However, it is challenging to encourage students to participate voluntarily in active studying processes for constructing their own meaning of knowledge for two reasons~\cite{krahenbuhl2016student}. First, as students have little knowledge of history and are not familiar with historical events, they often fail to find appropriate learning materials by themselves. Second, the sheer amount of information on history available for studying implies both metacognitive and analytical overloads during explorations of historical events. Without proper guidance, it is hard for students to perform the tasks required for active history studying based on constructivism theories. For example, studying history involves several tasks of searching for facts and finding relationships among the figures and countries in the events. Students are also required to investigate changes in regions according to time and to elucidate the possible causes of such changes. If the events are intertwined with each other across locations, or if they do not seem related to each other, students can easily become discouraged from completing the activities without effective assistance and guidance~\cite{poitras2014developing}. 
As such, there is a need to help students who want to study history actively in classes.

In this work, we introduce HisVA, a visual analytics system to augment pedagogical practices in flipped history classrooms by enabling students to explore historical events in a more self-directed way.
To design HisVA, we have collaborated with three domain experts in history and education and extracted system requirements. 
HisVA employs coordinated multiple views (CMVs)~\cite{roberts2007state} to provide an effective event exploration space, where users can explore desired events from a collection of relevant Wikipedia articles extracted for the purpose of the exploration.
The system also recommends highly related incidents from selected events to allow users to generate a narrative from a series of historical events. For evaluation, we provide two usage scenarios and conduct a user study and qualitative analysis of how HisVA is used. 
We also supply observations and the advantages of HisVA from the in-class deployment of HisVA for an assignment. 
The results indicate that HisVA is an effective visual tool providing enough temporal and spatial contexts for historical events and allowing users to efficiently find and understand historical events with users’ own questions produced using HisVA. 
The users most prefer the map view for their exploration. 
The instructors who reviewed the in-class assignment report that HisVA helps the class students to submit high-quality assignment results with diverse findings and questions not covered during lectures.

The main contributions of this work are as follows:
\begin{enumerate}
    \item Task analysis and derived system requirements for history education,
    \item Design of HisVA to provide an effective spatio-temporal exploration space, and
    \item Evaluation with two usage scenarios, a formal user study, qualitative analysis on exploration strategies, and in-class deployment.
\end{enumerate}

\section{Related Work}
\label{sec_related_work}
We introduce previous work in the perspectives of education methodology, tools, resources, and event visualization.

\subsection{Learner-centered History Education}
There are various types of instructional methods in traditional education (e.g., lectures, demonstrations, discussions, projects, experiments), and we group the methods into two categories based on the role of the class activities—instructor-and learner-centered education.
The instructor-centered education method focuses on an instructor’s opinion and is based on a strict use of curriculum and textbooks. 
This approach asks students to respond with the ``black and white'' answers~\cite{brooks1999search}. 
In contrast, the learner-centered education method~\cite{richardson2003constructivist} assumes that knowledge is not discovered, but rather constructed. 
Thus, it emphasizes that creating a learning environment~\cite{lebow1993constructivist, driscoll1994psychology} in which students can construct their own knowledge is important, because it is difficult for students to build their knowledge on their own, due to a lack of domain knowledge and cognitive overload~\cite{krahenbuhl2016student}. An increasing number of classes in the field of history education have begun to emphasize students’ active participation~\cite{mccarthy2000active, savich2008improving, clark2017surprise}. In these classes, students not only study major historical narratives, which are constructed by historians, but also work on their own to explore historical events and find their meanings. However, there is an obstacle in this approach, in that there is too much information for non-history major students, who have less domain knowledge, to organize properly. 
Thus, it becomes more important to offer students a systematic guide and relevant material that make it possible for them to study by themselves~\cite{poitras2014developing}.

\subsection{Tools for History Education}
A variety of tools support studying history. One common feature of the existing tools is that they focus on providing students with historical contents~\cite{cabiness2013integrating}. Another kind of tool aims to support students in their efforts to study history more actively~\cite{boadu2014examination}. Although these tools are intended to encourage learners’ active learning, oftentimes learners are not systematically and effectively guided to investigate history. There are also visualization systems designed to assist users in studying history (e.g.,~\cite{Itoh12, Cho16}). VAiRoma~\cite{Cho16} is the most recently developed visual tool with this purpose in mind. It provides geographical, trend, and topical information by using CMVs~\cite{roberts2007state} of maps, charts, and topic views, aiming to help users study effectively about Roman history. 
We note that Firat and Laramee provide an excellent survey on visualization systems for various education areas (e.g., engineering, medicine)~\cite{Firat18}.
Compared to the existing work, HisVA provides ample guidance for students’ spatiotemporal event exploration with its map and event views, along with topics computed from the Wikipedia corpus. In addition, HisVA suggests important events by computing document similarities and page-rank, to help users narrow down their navigation efforts effectively.

\begin{figure*}[t]
  \includegraphics[width=1\textwidth]{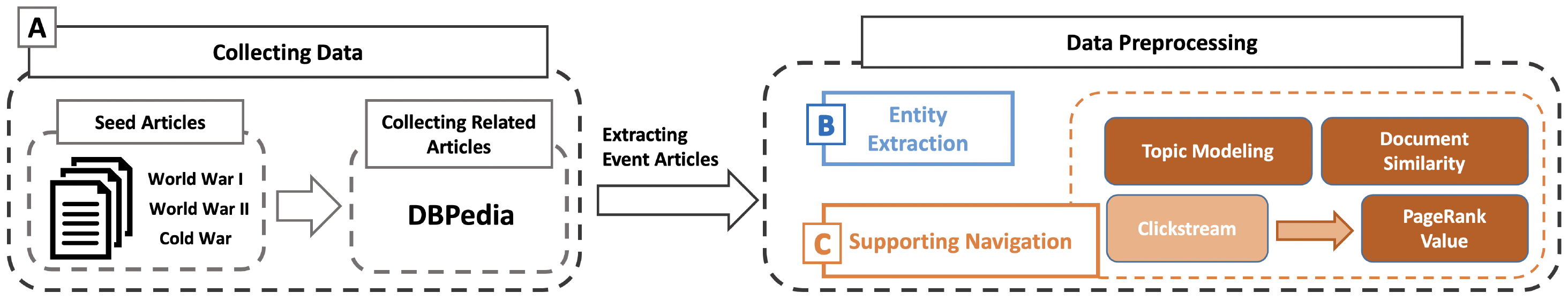}
  \caption{Data pre-processing pipeline: A) Collecting 3,019 historical events based on seed articles (e.g., ``World War I''), B) Extracting date and location information from the collected events, C) Computing for reducing the exploration space (e.g., topic modeling).}
  \label{fig_dataprocessing}
  \vspace{-0.5cm}
\end{figure*}

\subsection{Wikipedia and History Articles}
\label{sec_rw_wiki}
Wikipedia is a free online encyclopedia that anyone can read and edit. 
Thanks to this openness, the size of Wikipedia quickly grows, which attracts researchers in various domains.
But the openness also brings issues on accuracy, validity, and adequate coverage of Wikipedia's data. 
To investigate these issues, much research has been conducted comparing Wikipedia to other resources (e.g., encyclopedias) which are written by authoritative experts in multiple domains, including history~\cite{Rector08, Samoilenko17, Samoilenko18}.
Rector~\cite{Rector08} compares Wikipedia articles on history against Britannica, the Dictionary of American History, National Biography Online, and report that  Wikipedia's accuracy rate is 80\%, while other resources generally reflect a 95\% accuracy. 
Samoilenko et al.~\cite{Samoilenko17, Samoilenko18} compare Wikipedia to Encyclopedia Britannica, in terms of topic scopes and coverages.
Their results indicate that historical articles in Wikipedia cover more recent events (especially from the 19$^{th}$ century).
Additionally, they concluded that Wikipedia is more readable than Britannica.

Clickstream data for Wikipedia is a network that records how users navigate Wikipedia. 
The clickstream data consists of origin and destination pairs with numbers that represent how many times users use each pair during navigation. 
A great deal of research has been conducted with the clickstream data for social science research.
Dimitrov et al.~\cite{Dimitrov17} show that readers can navigate the Wikipedia articles to investigate articles that are semantically linked. 
Schwarzer et al.~\cite{Schwarzer16} compare the effectiveness of citation-based and similarity-based recommendation approaches for users' navigation.
Their results indicate that similarity-based recommendations help users identify articles that are semantically linked, while those based on citations can better assist users in finding subject-related information. 
Inspired by this result, we employ recommendation approaches in this work to better support users’ navigation of historical articles.

\subsection{Event Visualization}
There is a large body of event visualization techniques or systems, as events can be defined, interpreted, or focused differently.
Conventionally, the goal of research on events is to explore the frequency, causality, or combinations of the sequences produced by events. 
In this case, simple visual representations (e.g., rectangles~\cite{Nguyen18}) are utilized that are arranged to allow an overview or side-by-side comparisons.
When the sequences are complicated, the size of each visual representation becomes smaller and similar to pixel visualizations~\cite{Ko12}.  
Examples include the visualizations of user interaction sequences collected from web logs (e.g., ~\cite{Liu17b}) and electronic health records (e.g.,~\cite{Monroe13}).

As event data become more diverse, new event analysis tasks emerge that require more than pattern-matching techniques based on frequency and combinations of events.
Events may occur with multiple attributes, any of which may be crucial for analyzing them.
For example, Crouser et al. present a visual interface for analyzing search events extracted from news data.
Dou et al.~\cite{Dou12} develop LeadLine, which automatically detects events from news and social media posts and allows users to investigate them. 
Compared to LeadLine,
HisVA has a strength in supporting users to follow the chronological flow of events and make assumptions regarding the cause and effect of these events, whereas LeadLine allows users to analyze events from the “4Ws'' perspective (who, what, when, and where).
The two tools’ visualization design considerations are also different, which leads to different outcomes, such as the topic- and importance-based event view in HisVA.

In the education and learning analytics domain, event visualizations have been utilized to visualize a series of user interactions to explore student behavior and learning patterns. 
For example, Shi et al.~\cite{Shi15} design VisMOOC to help instructors and educational analysts understand students' online learning behaviors, presenting sequences of interactions with videos, such as play and pause. 
Chen et al.~\cite{chen2018viseq} introduce ViSeq that helps users identify different learner groups based on learning behaviors. 
Our tool is different from these previous work in the education and learning analytics domain in that we aim at directly helping students in their studying, whereas the previous work focused on log analysis.

\section{Task Analysis}
\label{task_analysis}
To design a visualization tool for studying history, we, as a team of multidisciplinary researchers composed of experts and instructors in computer science, history, cognitive science, and education, have met bi-weekly over 24 months to discuss history courses’ goals and procedures, and to uncover the difficulties that students face in history courses. 
Three of us have been developing and teaching history courses at UNIST for the past five years. 
Among us, two hold doctoral degrees, in history (E1) and education (E2), and one has a master’s degree in history. Next, we describe how the history courses at UNIST have been operated.

At first, the history courses have been run based on the flipped learning model~\cite{Bergmann12, O15}, which emphasizes active student participation in learning activities. 
Using the flipped learning model encourages students to form their own historical perspectives voluntarily, beyond simple knowledge acquisition from textbooks. 
In the flipped learning model, students are required to play an active and self-regulated role, 
while an instructor acts as facilitator rather than knowledge deliverer. 
As such, the interaction between an instructor and students, as well as among students, is strongly emphasized~\cite{Bergmann12, Kim14}.

\begin{figure*}[t]
  \includegraphics[width=1\textwidth]{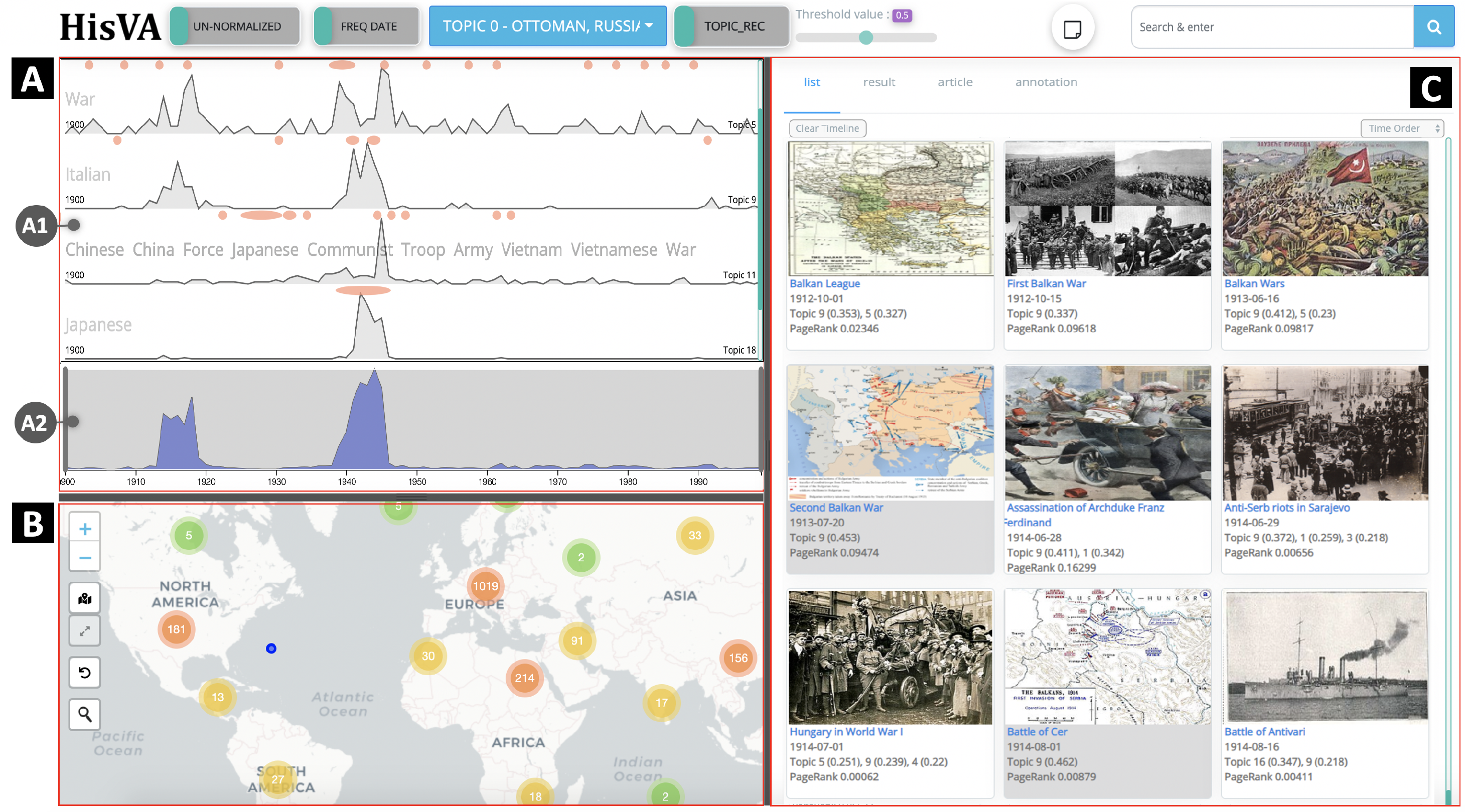}
  \caption{An overview of HisVA. Users can gain temporal, frequency, related topic, and important information of events in the event view (A), spatial features of events in the map view (B), and auxiliary information in the resource view (C).}
  \label{fig_teaser}
  \vspace{-0.5cm}
\end{figure*}

The history course based on flipped learning at UNIST has five main cycles, each lasting two weeks with four classes. For example, the course, ``History of Modern World,'' in the 2019 fall semester consisted of five cycles with five topics—``Imperialism and Colonialism,'' ``World War I and Postwar Changes,'' ``Interwar Era and World War II,'' ``the Cold War,'' and ``the Global Cold War.''
For each cycle, an instructor provides assignments and guidance to students to help them better understand each topic. 

Before the first class in each cycle, the instructor provides information on the topic, asks students to read a textbook chapter related to the topic, and prepares an in-class discussion and quiz. 
In the first class of each cycle, the instructor gives a lecture, providing an overall picture of the topic. In the next class, a group activity is performed in which group members are asked to carry out a task with materials on the given topic. In the third class, the instructor gives an extended lecture, introducing controversial points on the topic. In the last class, students share their thoughts on the topic in groups, and each group leader summarizes their thoughts and reports the results. 
The instructor moderates group discussions in each class, providing guidance and feedback on the activities. When each cycle is over, students are required to write an individual essay about what they have studied on the topic.

During our discussions, E1, E2 and E3 expressed that flipped learning is effective, but there are challenges for teaching history effectively. First, there is much content that students in the course need to review (i.e., large exploration space). Examples include not only the textbooks but also online materials that are returned by keyword searches. Thus, every semester, many students ask how to efficiently study history, and a few will fail in the end, as they are often lost during lectures and cannot efficiently review the many events in diverse regions and time ranges. These phenomena are common in education; they’re called disorientation~\cite{William01, Paulo99}, according to E2.

Second, there is an imbalance in studying history with existing courses and materials. Many universities offer a course called ``Western civilization,'' but rarely do universities offer courses in which students can study the history of other than Western regions and form a global, balanced perspective on history. The existing materials (e.g., textbooks) tend to place more importance on Western European history, although other regions also have crucial events in their history. For example, China, Indonesia, Iran, Ethiopia, Angola, Cuba, and Nicaragua played important roles in the Cold War~\cite{Westad05}, but the events related to the United States and the Soviet Union are those mainly highlighted in lectures. The materials are already selected, compiled, structured, and edited by historians, so relying on them could lead students to accept others’ interpretations and fixed narratives regarding history without criticism. This may constrain students' creativity and prevent the creation of critical questions on global history.

Third, helping students form their own balanced perspective is one of the courses’ goals. 
We believe that such a perspective can be instilled by the practice of seeking links between events that are seemingly unrelated. With this in mind, an instructor gives assignments in which students can discover new facts and relationships among other events and countries from a global perspective. But in running history courses over the semesters, they have found that such methods based on the assignments are not as effective as they expected, because most students submit similar interpretations on a given topic by reading similar articles returned in online searches. Conversely, when the materials from online searches are sparse or not detailed, the quality of the assignment results are often poor. Even when sufficient materials exist, it is still hard for some students to link events that are seemingly disparate. Finally, E1, E2 and E3 stress that they do not find tools that can resolve these issues.

From the discussions, we derived the following requirements for a visualization system designed for studying history:
\begin{enumerate}[(R1)]
\item Presents a reduced but effective exploration space; 
\item Supplies spatial and temporal contexts of historical events;
\item Assists in the acquisition of information for investigation of historical events; and
\item Promotes linking diverse events and finding relationships among them.
\end{enumerate}

We initially thought that the derived requirements involving existing spatiotemporal tasks~\cite{andrienko2010space}  could be supported by previous visualization systems. Thus, we used VAiRoma~\cite{Cho16} for a couple of history class assignments to identify the essential features to fulfill the requirements. Although VAiRoma provides features that partially support R1 and R3, we found some missing features for other requirements, because it was developed with a different motivation and target user in mind. For example, VAiRoma supports R4 by using CMVs, but it was difficult to find appropriate information from the large set of articles. Therefore, we decided to design and develop a new system to better support all of our present requirements.

\section{Visual Analytics System for History Education}
\label{sec_system}
In this section, we introduce HisVA, a visual analytics system for visualizing and navigating historical events from text collections. 
Note that any text collections could be used for HisVA; however, in this work, we focused on Wikipedia articles for the main topics (World War I and II, and Cold War) of our history class to illustrate the functionality of the system.
Next, we describe our data pre-processing pipeline and then present HisVA, a visual event exploration system for studying history, as shown in Fig.~\ref{fig_teaser}.

\subsection{Data Pre-processing}
\label{DATA_ORGANIZATION}

The first step of the data preparation methodology is to select a set of Wikipedia articles of interest to be used as a starting point. These articles are used as seeds to find related articles from DBpedia~\cite{Bizer09}, forming semantic relationships between Wikipedia articles. All articles that share the same subject categories as the seed articles are then extracted from DBPedia. From these, the ones marked as an ``event'' in the DBpedia ontology type are selected for exploration in HisVA. 
In this study, the seed articles used are ``World War I~\cite{Wiki_WW1},'' ``World War II~\cite{Wiki_WW2},'' and ``Cold War~\cite{Wiki_COLDWAR}'' as shown in Fig.~\ref{fig_dataprocessing}. 
We initially collected 5,467 Wikipedia articles, of which 3,019 marked as ``events'' were used.

Next, we extracted date and location entities from each article and performed pre-processing to provide users with the spatial and temporal contexts of an event (R2). For the extraction, we used the 7-class model Stanford Named Entity Recognizer (SNER)~\cite{Finkel05}, trained on the Message Understanding Conference (MUC) 6 and 7~\cite{grishman1996message} training data sets.
We then counted the number of dates and locations associated with each event and set those with the most frequently shown information as the representative temporal and spatial information of each event. For example, ``Germany'' and ``March 1945'' are mentioned 98 and 6 times, respectively, in the article ``World War II,'' so we used them to represent the location and date of ``World War II.'' 
We extracted geocoordinates (i.e., latitude and longitude) of the locations using Geopy~\cite{Geopy06}.

To provide an overview of the historical article collection (R1), we used the topics produced by topic modeling algorithms. To find an appropriate model, we tested topic modeling methods with various topic counts (e.g., 10–50 topics), including LSI (Latent Semantic Indexing)~\cite{hofmann1999probabilistic}, HDP (Hierarchical Dirichlet Process)~\cite{teh2005sharing}, LDA (Latent Dirichlet Allocation)~\cite{blei2003latent}, and LDA Mallet~\cite{McCallum02}. After reviewing the modeling results, we decided to use LDA Mallet with 20 topics, due to its highest coherence score (0.47), which measures the similarity of the words in each topic~\cite{roder2015exploring}. We used the topic modeling algorithms implemented in Gensim~\cite{rehurek_lrec} and the coherence model to measure the coherence score based on normalized point-wise mutual information (NPMI) and the cosine similarity (called coherence  C\textsubscript{V})~\cite{roder2015exploring}.

We used two methods for recommending important articles to help users begin their explorations (R3). The first is topic contribution, which is computed by a topic modeling algorithm and indicates how much an article contributes to forming each topic. As a second method, we use the articles’ centrality (i.e., page-rank) in a network, computed from the Wikipedia clickstream. The page-rank value of an article indicates its popularity, because page-rank values are proportional to the article’s number of views. HisVA provides a toggle button for the selection between contribution-based (TOPIC\_REC) and popularity-based recommendations, as shown at the top of Fig.~\ref{fig_teaser}. In addition to the toggle, there is the slider bar, which can be used to adjust threshold values. If an article’s contribution or popularity value is greater than the threshold value, it is recommended as an important article.

\subsection{Visual Interface} 
HisVA consists of three main views—event, map, and resource views—as shown in Fig.~\ref{fig_teaser}. 
We use CMVs~\cite{roberts2007state} to provide users with an interactive learning environment~\cite{Firat18} to help them continue acquiring knowledge by themselves in any of the views (i.e., self-directed learning)~\cite{Whitelaw15, Mayr16}. 
CMVs can also facilitate linking diverse events and finding relationships among them from different perspectives (R4). 
A keyword-based article search system can be an alternative, but such systems can be difficult for casual users to navigate~\cite{Windhager18}. 
We develop HisVA with D3.js \cite{D3js}, Leaflet \cite{Leafletjs}, and Flask~\cite{Flask10} and use Leaflet's marker cluster to aggregate the markers on the map~\cite{Leaflet_MarkerCluster}.

We first explain the filter view (Fig.~\ref{fig_teaser}, top) to provide a better understanding of how to utilize the three main views. The filter view has three toggles (default: off, gray color), a drop-down menu, a slider bar, and a search-keyword input interface. The toggle for normalization allows users to normalize the y-axes of the line charts in the event view (Fig.~\ref{fig_event_view} A and B).
The toggle for computing date frequency allows users to choose whether to use all the dates associated with an event (Fig.~\ref{fig_event_view} D) or only one representative date (Fig.~\ref{fig_event_view} C) when computing the number of events in each year of the event chart (Section~\ref{sec_event_view}).
We further describe the normalization and date frequency toggles in Section 4.2.1. 
Users can select the system's recommendation option, using the toggle to choose between a topic recommendation (``TOPIC\_REC'') or popularity recommendation (``POPULAR\_REC''). 
To help users understand the options, a tooltip is shown to describe the topic and popularity recommendations whenever a user hovers the mouse over the options. 

Users can set a threshold value for article topic weights (i.e., how much an article contributes to extracting a topic during the topic-modeling process) and page-rank value with the slider bar, make notes, (\faIcon[regular]{sticky-note}), and search keywords. 
Next, we describe the three main views for visual historical event exploration—event, map, and resource views (Fig.~\ref{fig_teaser} A, B, and C).

\subsubsection{Event View}
\label{sec_event_view}

\begin{figure}
    \includegraphics[width=0.5\textwidth]{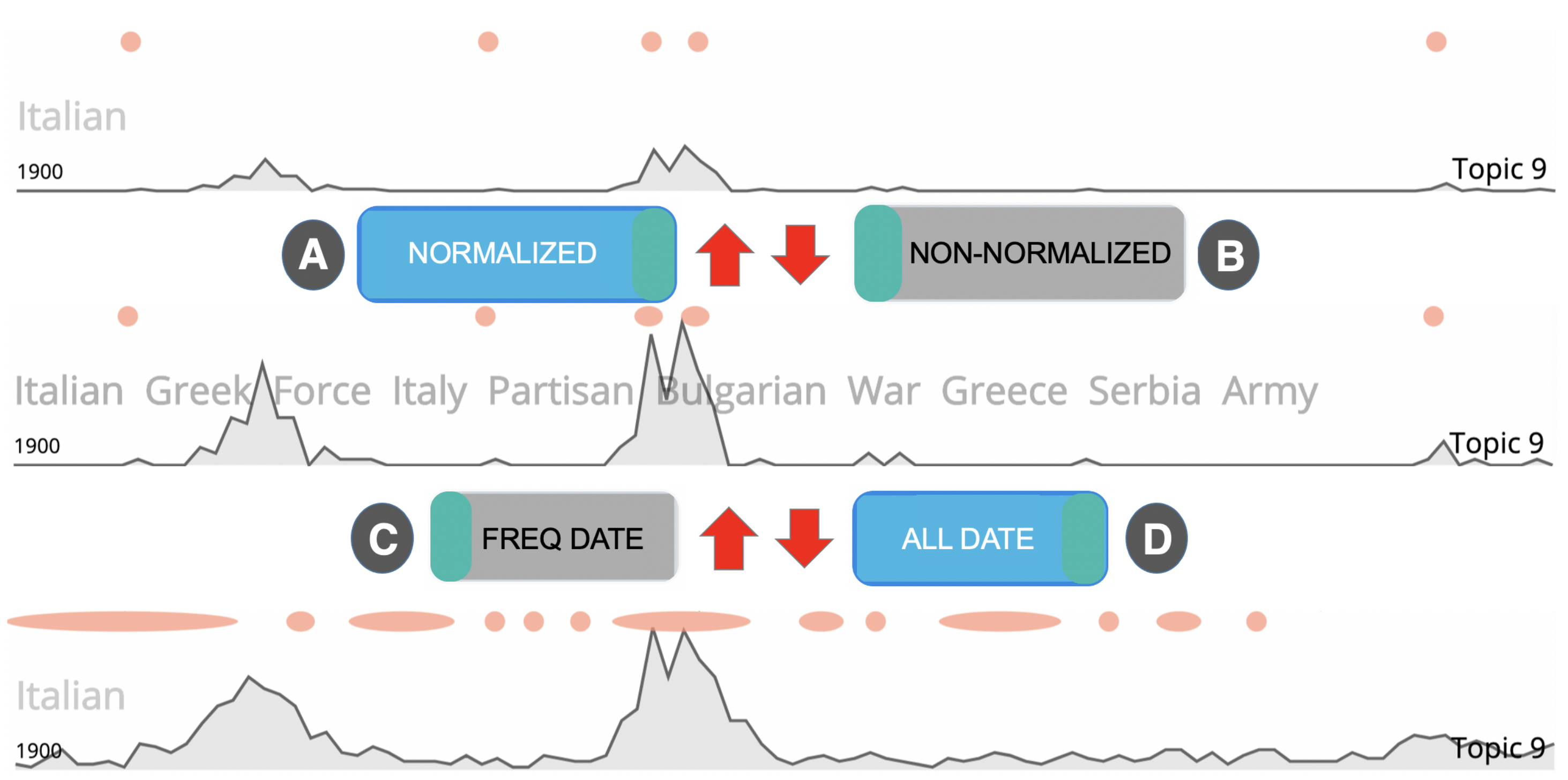}
     \caption {
     Each event chart presents topic, topic keywords, important events (pink dots), and a line chart for representing the number of events associated with the corresponding topic. A basic chart for the topic ``Italian'' (middle), a normalized version (top), and a chart which the ``all dates'' option applied (bottom).
     } 
    \label{fig_event_view}
    \vspace{-0.4cm}
\end{figure}

The event view (Fig.~\ref{fig_teaser} A) presents multiple event charts and a summary chart. 
We place twenty event charts that are constructed based on computed topics (R1) and time (R2) information.
The charts are initially ordered as produced by the modeling algorithm, but users can scroll the view, interactively adjust the chart order, and hide and show any event chart they choose. 
Each chart shows 10 topic keywords, which are ordered by their contribution to each topic (Fig.~\ref{fig_event_view} middle). 

\textbf{Summary Chart:} The summary chart (a line chart at the bottom of the event view, Fig.~\ref{fig_teaser} A2) shows the aggregated number of the articles by time to let users see the number of events across time and important events in specific time ranges (R1). 
For example, the summary chart in Fig.~\ref{fig_teaser} shows two peaks from 1910 to 1920 and from 1940 to 1945. The peaks mean that, given the document corpus, some events in the two time ranges have the greatest number of articles related to World War I and World War II. 
In addition, the summary chart provides two vertical gray bars at each side which users use for filtering time ranges (R1, R2). 
If the bars are at each end, the entire data set is used for computing the number of events by time.

\textbf{Event Chart:} Each event chart (e.g., Fig.~\ref{fig_teaser} A1) has a line chart for presenting the number of events associated with the topic, topic number, and representative topic keyword. 
The x-axis is the time that is the same as that in the summary chart, while the y-axis presents the number of events associated with the time.

When the normalization option (Fig.~\ref{fig_event_view} A) is on, the event’s article count (i.e., y-axis) is divided by the largest document count. 
We provide the normalization option as the number of articles for each topic is not evenly distributed. 
Each chart is non-normalized by default because it is hard to notice changes (e.g., peaks, dips) when a topic has fewer events than others (e.g., different heights on the same number of events in Fig.~\ref{fig_event_view}, top (normalized) and middle charts).
In the state of FREQ DATE option (Fig.~\ref{fig_event_view} C), HisVA maps an event with only one date, the most frequently appearing in the article. 
For example, the event World War II is only associated with 1945-03-01, which is the most frequent date for the World War II document. 
When the ALL DATE option (Fig.~\ref{fig_event_view} D) is on, the World War II event is linked to several dates (e.g., 1935-10-03, 1941-06-22) that appear in the article.

\begin{figure}
    \includegraphics[width=0.5\textwidth]{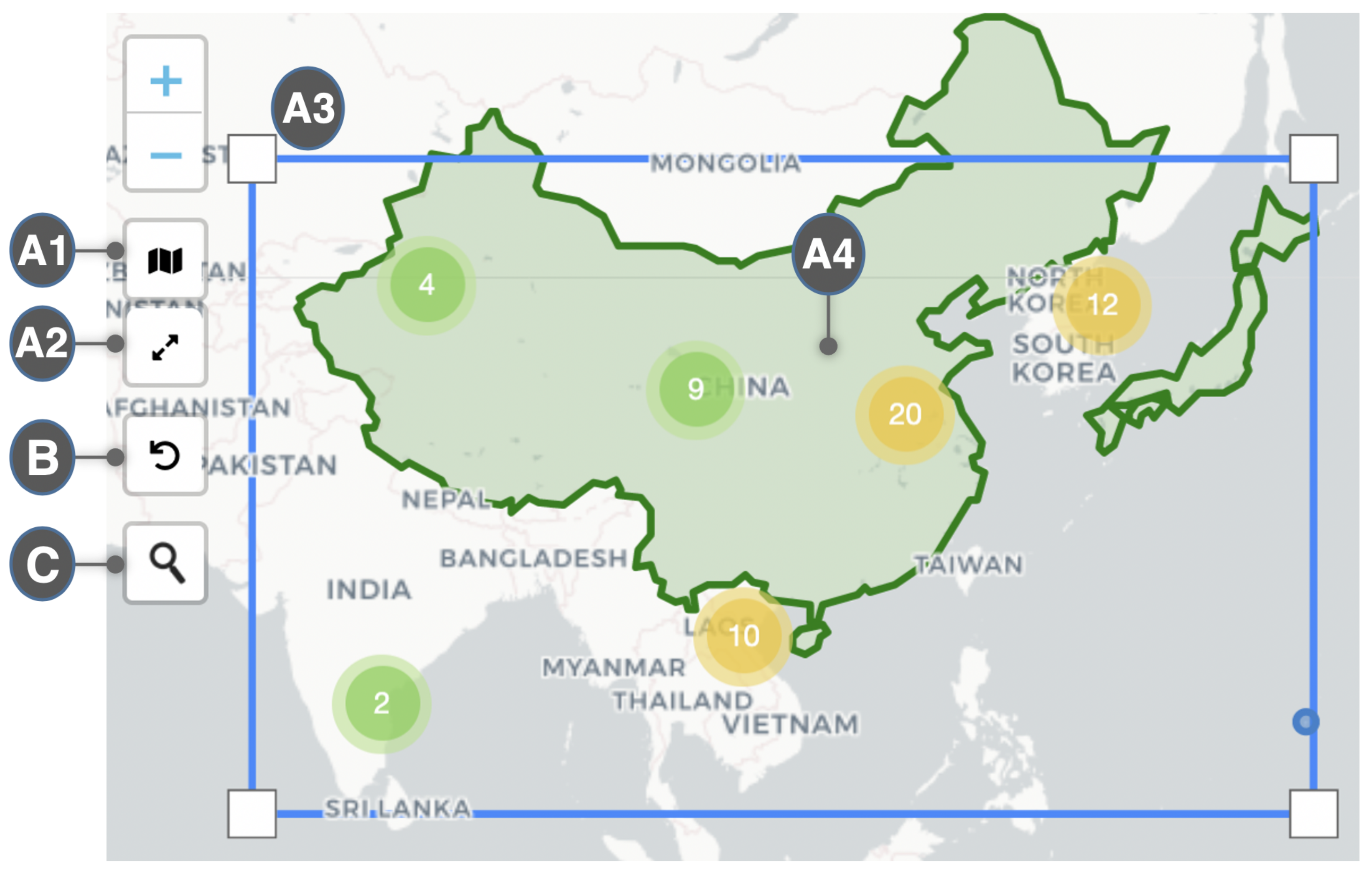}
     \caption{
     Each aggregated circle means the number of events that occurred in each region. Users can filter events by drawing a box (A3), selecting countries, or searching country names (C).
     }
    \label{fig_map_view}
    \vspace{-0.5cm}
\end{figure}

\textbf{Displaying Important Articles:}
The events with higher importance weights (i.e., topic contribution or page-rank scores) than the threshold value are presented as a pink dot at the top of the event chart (R1). 
When a series of adjacent dots occlude each other, they are aggregated into a wide dot (i.e., ellipse).
For example, the third event chart in Fig.~\ref{fig_event_view} shows that the topic of the chart is ``Italian,'' and there are seven important events and six wide dots which contain more than two important events. 
Note that the number of articles in a year is not related to the computed importance of events, although there is high chance that important articles will be found around peaks (e.g., the World Wars).
When users hover the cursor on a dot, a tooltip pops up, containing the article thumbnail associated with the dot, as shown in Fig.~\ref{fig_case_study1} B and C.
Users can click a thumbnail to read its Wikipedia article in the article view (Section~\ref{RESULT_VIEW}).

In designing the visual cue for important articles, we intend that the representation should be easy to understand when used for not only individual articles but also multiple aggregated articles. 
We also consider the need for easy interaction of the representations with users. 
With these considerations, we review several symbols (e.g., a dot, triangle, rectangle, line), as they are easy to study and perceive~\cite{ware2019information}, but find that not many symbols satisfy our considerations. 
For example, we exclude triangles because when a triangle is placed in the middle of two topic charts in the topic view, a vertex point can be read as a direction (e.g., upward direction). 
We also rule out lines as it is not easy to hover and click them. 
A square is not suitable, as it cannot be resized uniformly, which is a weakness for aggregated articles. 
In the end, we decide to use dots for individual important articles and ellipses for aggregated articles. 
A rectangle and bordered rectangle can be used for the same purpose; however, it is possible that the start and end points of two adjacent and closely-placed rectangles might not be easily distinguished~\cite{bertin1983semiology}. To resolve this issue, Lee et al.~\cite{lee2019visual} suggest to slightly bend each end side of a rectangle. 
We suppose that ellipses would have the same effect as bent rectangles.

\subsubsection{Map View}
\label{MAP_VIEW}

The map view (Fig.~\ref{fig_map_view}) displays locations associated with historical events. The basic unit to represent events is a cluster marker with a total event number in the region. 
Users can change the zoom level of the current map extension by using the magnification icons (Fig.~\ref{fig_map_view}
\raisebox{-0.4ex}{\includegraphics[width=0.03\linewidth]{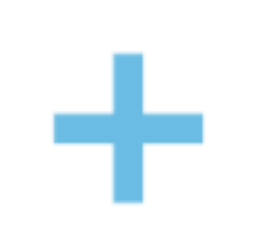}}, \raisebox{-0.4ex}{\includegraphics[width=0.03\linewidth]{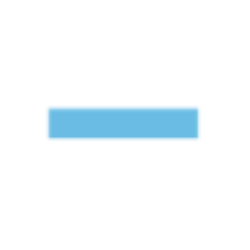}}), the mouse wheel, or by clicking any of the clusters. 
When the map is zoomed out, adjacent clusters are aggregated into a bigger one, and when it is zoomed in, cluster markers split into multiple markers, which are relocated based on their own regions of events. 
When the zoom level reaches its maximum, each marker is surrounded by circles, each of which is mapped to an event. If the mapped event is clicked, it is searched and highlighted in the list view (Fig.~\ref{fig_teaser} C) for reading. 
When the cursor hovers over a circle, a tooltip pops up on the circle showing the title and date of the associated event.

Users can investigate regions of interest in three ways. 
First, they can draw a box to specify a region of interest on the map (Fig.~\ref{fig_map_view} A3) after clicking Fig.~\ref{fig_map_view} A1. 
When this interaction occurs, the events in the box are shown in the list view for reading. 
Users can remove a box by clicking (Fig.~\ref{fig_map_view} A2). 
Second, users can also directly click the countries one by one (e.g., China and Japan in Fig.~\ref{fig_map_view} A4). 
When a region or country is specified on the map, event charts present a red rectangle to indicate the time span of the events in the specified region (Fig.~\ref{fig_case_study1} B and C).
Lastly, in the map view, users can search regions by name (Fig.~\ref{fig_map_view} C). 
They can also refresh the map (Fig.~\ref{fig_map_view} B) to deselect all regions.

\subsubsection{Resource View} 
\label{RESULT_VIEW}

To help users efficiently investigate historical events with detailed information (R3), HisVA provides a resource view (Fig.~\ref{fig_teaser} C) comprising four sub-views—list, result, article, and annotation views. 
The list view showcases historical events, allowing users to sequentially access a series of events, along with meta information, such as a thumbnail, event date, topic number associated with the article, topic weight, and page-rank value, as shown in Fig.~\ref{fig_teaser} C. 
For efficient exploration, the list view allows users to sort the events by date, important weight, and topic (R1). Events with a importance weight higher than the threshold value are highlighted in gray (e.g., ``Second Balkan War'' in Fig.~\ref{fig_teaser} C). 

\begin{figure}
    \includegraphics[width=\columnwidth]{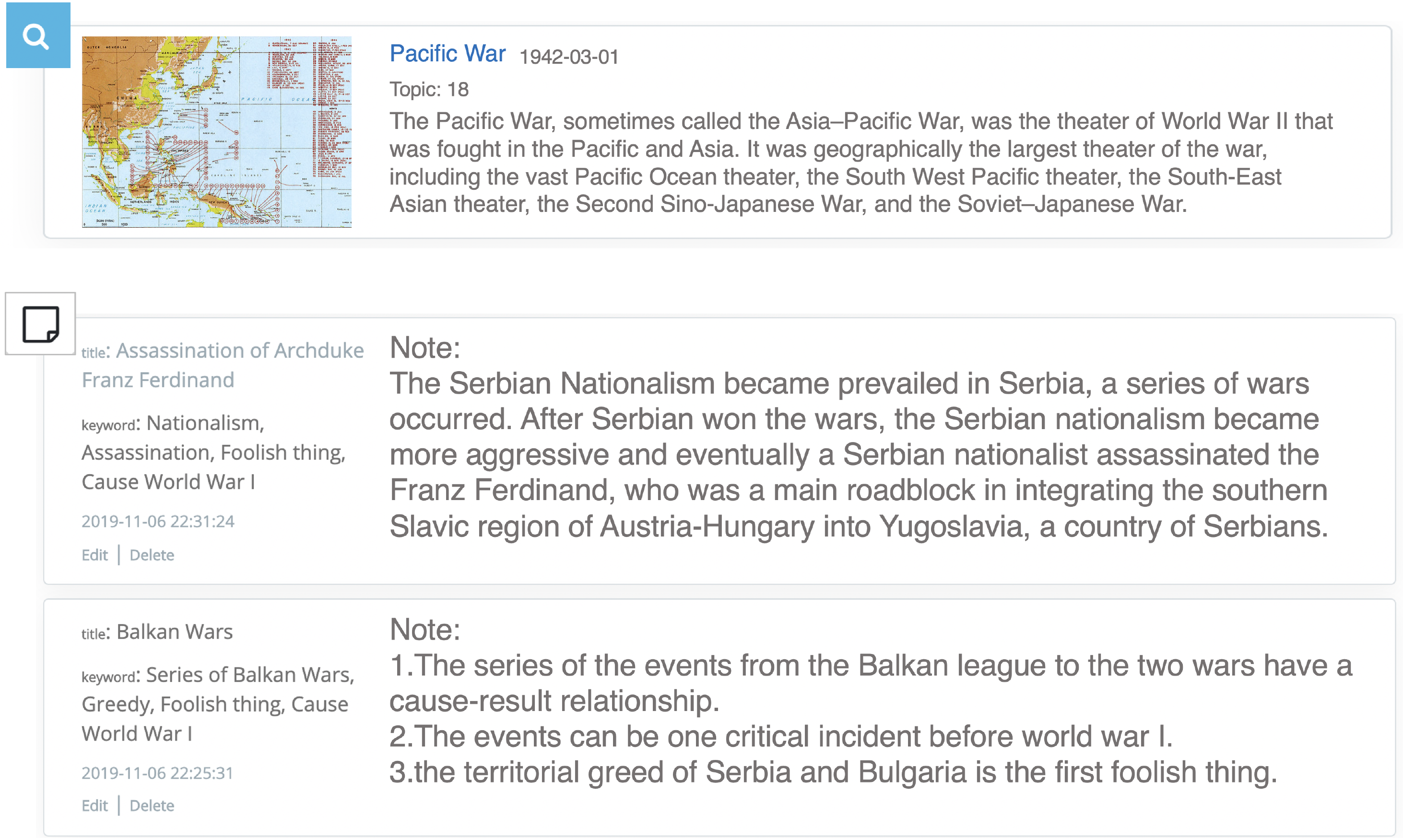}
     \caption{
     Users can search events with keywords (top) and make notes to record their thoughts during exploration (bottom).
     }
   \label{fig_resource_view}
   \vspace{-0.5cm}
\end{figure}

The result view lists articles in response to the user’s search in the search bar (Fig.~\ref{fig_resource_view} top \faIcon{search}). 
We present articles with the linear structure for easy content navigation in studying~\cite{Liang09}. 
When users click on an event, a new window pops up and presents the clicked event with other events that are most relevant to the clicked event (R4) (Fig.~\ref{fig_case_study2} C). We use the similarity of articles and transition counts as a relevance score. We compute the cosine similarity using the doc2vec model in Genism~\cite{rehurek_lrec} and the transition counts in the clickstream data to measure articles’ association ~\cite{Dimitrov17}. The transition counts indicate how many Wikipedia users make a transition from one article to another article using the internal links in Wikipedia articles. Users can choose which method to use for the event recommendation between the cosine similarity and transition count, toggling the option between ``TOPIC\_REC'' and ``POPULAR\_REC'' in the filter view.
When an article is selected in this view, it is shown in the article view for reading. Lastly, the note view presents a list of notes that users create in a template for a related article title, keywords, and creation date, as shown in Fig.~\ref{fig_resource_view}, bottom \faIcon[regular]{sticky-note}.
By clicking on the title, users can refer back to the original article from the annotation.

\subsubsection{Coordinated Multiple Views}
To facilitate the task of linking events and finding relationships among them (R4), HisVA utilizes CMVs~\cite{roberts2007state} with quick and easy interactions. For example, when a time range is set in the summary chart (Fig.~\ref{fig_teaser} A2), the historical events in the user-specified time range are presented in all map, article, and event views. When the event’s importance or frequency weight is changed due to user interactions with the filters (Fig.~\ref{fig_teaser} top), the visualization and articles in each view are also updated. 
In the same context, when users select regions by drawing a box or clicking countries in the map view, visualizations and articles in other views are updated accordingly. When users select an event from a cluster marker, the article view is auto-scrolled to present the part of the article associated with the selected event.
When users hover the mouse over an article in the list view, the center of the map view is moved to the location associated with the selected article and the cluster marker associated with the selected article changes to the spiral representation with the clicked article highlighted in red.
\section{Usage Scenarios}

\label{sec_case_study}
We provide below two usage scenarios that demonstrate how a user investigates relationships among events and performs event exploration with their own questions and motivations. 
We choose these as they are typically performed for essay assignments and in-class discussion. 
We assume that the students in the scenarios learned about World War I and II from lectures.

\subsection{Investigating Relationships of Historical Events}
In history courses, students are often given the task of finding relationships between events (e.g., cause and result). 
In this study, we show how a student uses HisVA to perform such a task in answering a sample question: \textit{``Bismarck famously said that a European war would start from some foolish things in the Balkans.
What \textbf{foolish things} happened in the region in the decade before World War I?''~\cite{Alpha_History}} 
It is directly related to contemporary history—one of the main topics in the history course of Fall 2019 at our university. 
The instructors confirmed that there are no direct online materials that provide well-structured and organized arguments regarding this question.

John, in a history course, is given an assignment to find the ``foolish things.'' 
At first, he searches the main question's keyword, ``foolish things, World War I'' on the internet. Reviewing the search results, he sees that some articles mention Wikipedia article titles. They seem to be related to the question, but none of them provides a direct answer to the question.

\begin{figure}[t]
\includegraphics[width=0.5\textwidth]{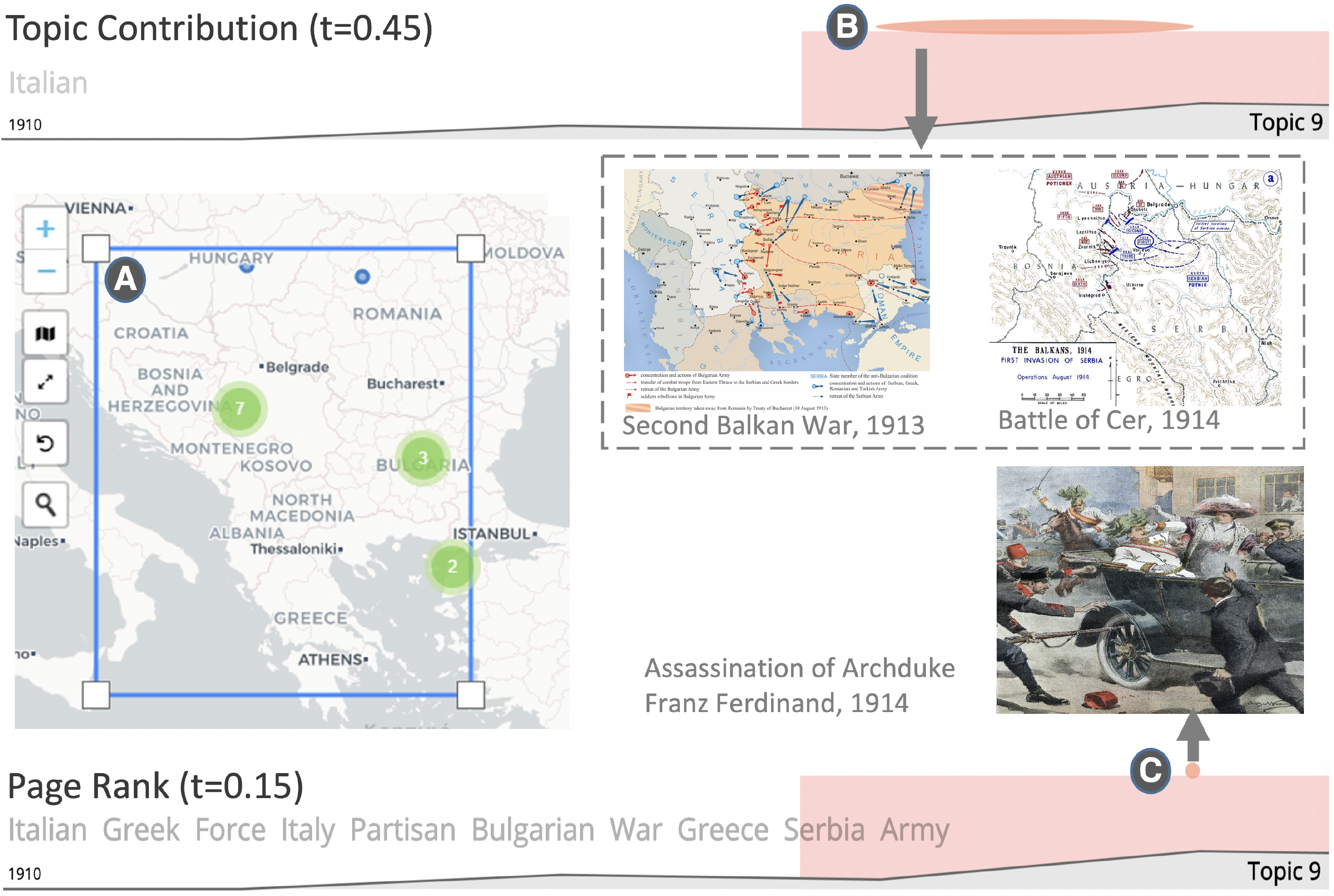}
\caption{
User-driven (e.g., (A) region of interest box) and data-driven (e.g., (B) recommendations, and (C) dots) exploration methods in HisVA.
}
\label{fig_case_study1}   
\vspace{-0.5cm}
\end{figure}

Next, he starts using HisVA. 
As he sees the temporal context from the description--``before World War I,'' he first sets the time range to the early 1910s, by using the vertical bars in Fig.~\ref{fig_teaser} A2.
To understand historical circumstances with a spatial context, he selects the Balkan areas on the map using the search function, which is given from the description and finds 3 markers that indicate 12 historical events in it (Fig.~\ref{fig_case_study1} A).
To filter by countries, he selects various countries on the map, finding the 12 events distributed in (Serbia: 4, Bulgaria: 3, Bosnia: 2, Romania:1, Montenegro:1, Hungary:1).

Reviewing the events in the list view, he sees that the events are mostly related to War (Topic 5), Greek (Topic 9), German (Topic 16), and Army (Topic 19). Thus, he filters the topics in the event view for further investigation by topic. As he adjusts the contribution option, a wide dot in the event chart (Fig.~\ref{fig_case_study1} B) for Topic 9 (keywords: Italian, Greek, (...), Serbia) captures his attention. As he hovers on the dot, HisVA pops up a tooltip, suggesting two articles--``Second Balkan War'' and ``Battle of Cer'' with event dates of 1913 and 1914. 
After he turns on the page-rank option and adjusts the threshold (0.16), an event captures his attention in the event chart with the word ``assassination,'' as shown in Fig.~\ref{fig_case_study1}. 
When he goes to the list view, all the suggested important events are shown as a result of spatiotemporal filtering. At this point, he thinks that investigating the events may help him answer the question.

As he focuses on the list view, he starts investigating how the events are related to each other and to the important events that he found in the event chart. Using the information in the pop-up view that recommends related articles, he finds that the articles relate mainly to three types of events in the region--1) Balkan war (``Balkan War,'' ``Second Balkan War'' in Fig.~\ref{fig_teaser}C), 2) Assassination (``Assassination of Archduke Franz Ferdinand'' in Fig.~\ref{fig_teaser}C), and Battle of Cer. 
He also notes that ``July Crisis'' is often recommended in the articles on Balkan Wars and Assassination, as it describes a series of intertwined diplomatic and military escalations, including the two wars and the assassination.

He starts reading the articles in the list view (Fig.~\ref{fig_teaser} C). In the first article--``Balkan League in 1912''—he finds that Greece, Montenegro, Bulgaria, and Serbia were allies in a fight against the Ottoman Empire. As the next article seems related to his task with the ``war'' and ``Balkan'' keywords, he reads the second article--``First Balkan War,'' which broke out 14 days after the date of the Balkan league event. 
The main story of the event is that the allies fought against the Ottoman Empire and almost won. 
However, a conflict arising due to territorial greed of Serbia in the first war led them into the ``Second Balkan War.''
In the second war (Fig.~\ref{fig_teaser} C, second row, first column), Bulgaria fought against Serbia and its allies (Romania, Greece, Montenegro, and the Ottoman Empire). He is interested in the Ottoman Empire’s support of Serbia, who defeated the Ottoman Empire in the previous war. 
The results of the second war were that Serbia won, Bulgaria blamed Serbia, and Albania became independent. 
Then he notes that 1) the succession of events from the Balkan League to the two wars has a cause–result relationship, 2) the events may be a critical incident before World War I, and 3) the territorial greed of Serbia and Bulgaria qualify as a foolish thing.

After reading the article on the assassination, he notes the following (Fig.~\ref{fig_resource_view} bottom):  Due to prevailing Serbian Nationalism, a series of wars occurred. After Serbia won the wars, Serbian Nationalism became more aggressive, and eventually, a Serbian Nationalist assassinated Franz Ferdinand, who was a major roadblock to integrating the southern Slavic region of Austria-Hungary into Yugoslavia, a country of Serbians. As the assassination is considered a direct cause of World War I, in which most European countries participated, he picks the assassination as the second foolish thing for the assignment.

\begin{figure}[t]
\includegraphics[width=0.5\textwidth]{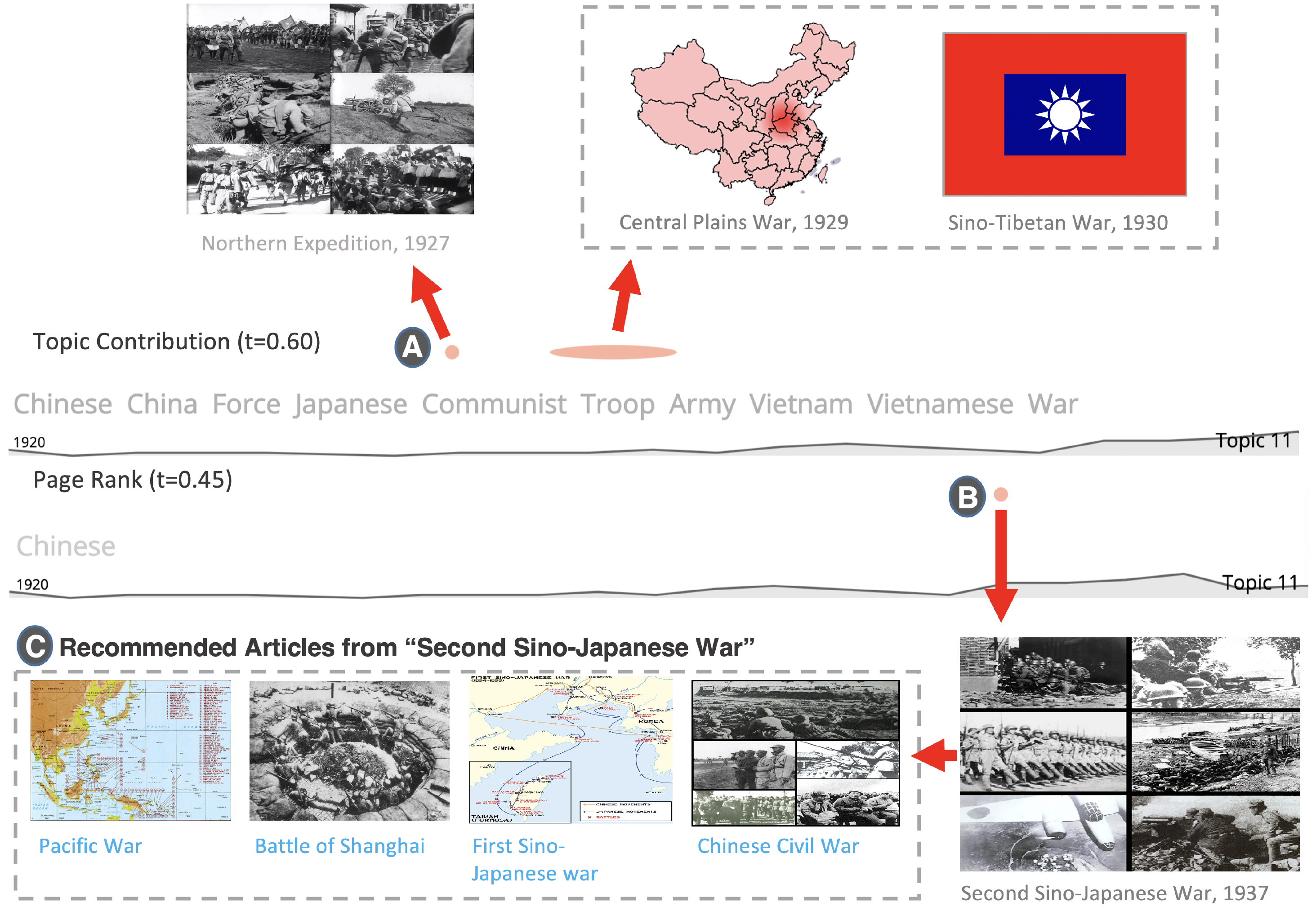}
\caption{Effective navigation with HisVA--(A) pink dots point out important events regarding topics; (B) articles recommended by other Wikipedia users.} 
\label{fig_case_study2}   
\vspace{-0.5cm}
\end{figure}

\subsection{Visual Exploration of Historical Events}
\label{sec_case_study_2}

In this section, we demonstrate that HisVA is effective in helping users explore events based on personal preference, knowledge, and perspective. Such exploration could improve user experience in studying and further motivate future self-directed learning~\cite{Hammond14}.

A student searches for historical events during World War II in preparation for a group discussion in a history course session. After logging into HisVA, she briefly reviews the information in each view in the default setting. During this review, she finds two attractive topic charts—Chinese (Topic 11) and Japanese (Topic 18)—in the topic view, as the countries are close to where she lives.
As she decides to focus on the two countries, she filters out others, leaving those two topics. The summary chart (Fig.~\ref{fig_teaser} A2) shows two peaks, indicating that the countries have been closely related between 1910--1920 and 1938--1945. 
As she knows that Japan engaged in World War II and colonized several countries in Southeast Asia at that time, she becomes curious about what events the countries were involved in together before World War II (i.e., before the second peak). First, she notices that the event chart for China (Topic 11) points out the important events with dots, such as ``Northern Expedition,'' ``Central Plains War,'' and ``Sino-Tibetan War,'' all of which were battles related to Japan, China, and other east Asian countries (Fig.~\ref{fig_case_study2}A). 
To further check which events other people were most interested in, she turns on the popularity recommendation option and sees one more event between Japan and China--``the Second Sino-Japanese War''(Fig.~\ref{fig_case_study2}B).

After exploring events with HisVA, she starts reading the articles in the list view. First, she reads ``the Second Sino-Japanese War'' (Fig. \ref{fig_case_study2}B) that describes how Japan initiated in 1937 and continued a war to colonize China until the end of World War II. Reading the article, she finds that Sino is China’s old name, and then poses a question on why Japan entered World War II during the war with China, as she recognizes with surprise that the Sino-Japanese war continued until the end of World War II. While exploring other articles to answer the question, she finds the article ``Pacific War'' because it is suggested by HisVA in the recommendation view, as shown in Fig.~\ref{fig_case_study2}C.

Reading the article, she understands that the relationship between Japan and the United States had been cordial but deteriorated when Japan invaded Manchuria in 1931 (``Japan’s invasion of Manchuria''). 
Despite warnings from Western countries (e.g., the United States, the United Kingdom, France), Japan continued its colonial expansion over the next decade, and finally, the United States stopped Japan’s oil imports in 1941 to quell Japan's expansionism. 
At this point, Japan knew that the relationship with the United States was severed and the lack of oil for the war with China would end with Japan losing the war. 
Japan finally decided to attack Hawaii to push the circumstance (``Japanese Attack on Pearl Harbor''). 
To prepare for the group discussion, she summarizes what she found from HisVA—how she started her exploration, posed a question during the exploration, investigated the relationships among the countries in the imperialism era, and came to understand some of the causes and another story behind World War II.

\section{User Study}
\label{sec_user_study}
To evaluate HisVA, we performed two user studies with two groups to measure the usefulness and generalizability of HisVA. First, we conducted a study with seven target users recruited from the history course at the university, followed by another study with 16 users from diverse backgrounds.

\subsection{User Study Task Design}
We designed two tasks for evaluation. Task 1 asked the participants to select three interesting events with HisVA and write down the reasons for their selection.
This type of task is assigned frequently in the course to encourage self-directed learning~\cite{Hammond14} in preparing for group discussions and activities in history courses.
Task 2 required the participants to find characteristics of the events in the 1910s in Africa using HisVA. 
The goal of Task 1 was to find out how HisVA supports users in exploring information and finding interesting events, while Task 2’s was to identify how HisVA helps users study World War I, one of the main topics of the history course in Fall 2019 at the university (Section~\ref{task_analysis}). 
As Task 2 involved Africa, which is less covered in internal history courses, it also allowed us to measure the effectiveness of HisVA when less familiar topics are given to students during the course.
  
\subsection{Participants and Procedure}
In Study 1, we recruited 7 participants by promoting our experiment in the history course (``History of Contemporary World'') of the Fall 2019 semester at our university. 
The participants were our target users, as they take the history course described in Section~\ref{task_analysis}.
The participants used a laptop (Intel i7, 2.9GHz, 16GB memory, Intel HD Graphics 630, 15-inch monitor) to have a similar environment to the one in the course. 
They were all males (average age: 22) studying various majors, including computer science, life science, and management.
Note that although the participants were recruited from the course, the results of the study did not affect participants' grades in the course, as no personally identifiable information was collected or sent to the course instructors. The course instructors were not involved in investigating participants' data.

Upon their arrival at the experiment room, we asked them to read and sign their consent to participate in the study. 
Next, we requested that they fill out a demographic questionnaire before beginning the training session in which they read the instructions, watched a 4-minute demo video, and used HisVA for about 15 minutes. After the training session, we asked the participants to perform Tasks 1 and 2 with HisVA.
We recorded the participants' activities on the screen and logged user interactions during the studies. 
After Task 1, the participants were allowed to take up to a 10-minute break. 
After Task 1, we asked them: 1) ``What function in HisVA (e.g., topic, summary charts, dots) mainly helped you find each event?'' and 2) ``What motivation did you have during your exploration with HisVA for each event (e.g., I traveled the country, so I wanted to know more about what happened in the country)?''
After Task 2, the participants completed a post-experiment questionnaire to rate the usefulness of HisVA and report their experience.  
The entire experiment took 100 minutes on average, including the break. We paid approximately \$17 as compensation for their participation. 

As we also wanted to find out whether HisVA is useful for students who do not have a  background in or knowledge of history, we conducted Study 2. In Study 2, we recruited 16 participants (6 males, 10 females, average age: 23) via flyers at our university. The tasks and procedures in the second study were much the same as those in the first, but for the second study, we used a desktop (Intel I7-6400, 3.4GHz, 16GB memory, GeForce GTX 970) with a 30-inch monitor (2560x1600) to simulate a conventional learning environment, like in a library or at home, as requested by the instructors. From the survey, we found that four participants had taken the history course and none had prior experience with visualization systems.

\subsection{Qualitative Analysis of How Students Use HisVA}
We report our qualitative analysis results based on interactions logs, recorded videos, and survey answers. 
Two authors coded the functions that were used for exploration and found 70 function use cases (a participant reported 4 events, so we had a total of 70 events in the dataset).
We chose to code primary functions, as the functions used could affect user experience in narrowing the large exploration space. 
We also thought that observing the most frequently used functions can help us find which functions are effective from the user’s perspective.
The primary functions included for coding were the map view, event view, search, recommendations, and combinations of those.
The functions provided in the event view, such as topic keywords, event charts, and pink dots of important events, were also included individually, which led us to have 18 function categories. 
In the coding, we used keyword matching. For example, we linked ``I found many events in Europe on the map view [...]'' to the map view category.

We performed thematic coding~\cite{thomas2003general} with user survey results to categorize 70 user motivations, as the coding results can tell us what triggers the use of the functions in HisVA and what aspects can be further considered when designing visualization systems for education.
As a result of the coding, we grouped the 70 motivations into 16 categories—using HisVA (i.e., map, topic keyword, timeline, articles in list view), prior knowledge from other classes, personal experience (e.g., travel, discussions with friends), and 10 combination categories (e.g., map + prior knowledge).
Example motivations generated from using HisVA included, ``When I see two peaks in the timeline, I wondered what events happened at that time to produce such large numbers of articles,'' ``I saw the dots (i.e., important events) in a topic chart, so I first checked the events associated with the dots.''
Note that the motivation for using HisVA can be more than one, as users can formulate multiple questions from several functions provided in HisVA. 
When the coders had a conflict, the last author involved in the codification resolved the conflict (Cohen’s Kappa: 0.952 and 0.950 for the coded functions and motivations, respectively).
\\
\textbf{HisVA Helps Users Develop Various Exploration Strategies: }
We found that 65\% (45 of 70) of the events were explored using multiple views.
Twenty participants reported that they used both the map view and event charts in the event view as their main exploration method.
When using this strategy, they found regions of interest on the map and then observed the changes in the number of events in the clusters on the regions by adjusting time ranges in the summary chart. 
The opposite strategy was also observed. 
In this case, participants kept adjusting time ranges in the summary chart, panning and zooming the map, and watching how the numbers in the entire clusters on the map were affected. Once they found a region of interest, they zoomed in on it for further investigation.
P18 shared her experience of how she explored events with this strategy. As she set a time range and brushed Africa, she found three events (``Ekumeku Movement,'' ``Battle of Zanzibar,'' and ``Battle of Kakamas''), located in the west, east, and south of Africa, respectively. Then she wondered why the events stretched across Africa. After reading the articles, she understood that the events were related to two different countries who advanced into Africa for colonization (Great Britain into western Africa and Germany into southern Africa). She then surmised that Africa had become a battlefield between the imperialist countries seeking to extract more resources from Africa, and also between the imperialists and the African countries wanting their independence.

We found that 36\% (25 of 70) of the events were explored with only one view.
The map view was the most popular, helping participants find 16 events. 
The main process in using the map was that they decided on a region and checked the events around the region, performing zooming and panning. We attribute this popularity and interaction pattern to their greater familiarity and the uncomplicated visual representations (e.g., markers) and interactions (e.g., zooming and panning) with maps that have been adopted in many map-based applications (e.g., Yelp and Google maps). Many participants described in detail how and why they utilized the map for their exploration. 
For example, P1 commented, \textit{``I found an event—`German attacks on Nauru'—when I searched for events in the Pacific Ocean on the map. (...) I decided to read the article, as I wondered why Germany came to the Pacific Ocean.''} 
P2 reported a similar experience: \textit{I saw Egypt with many events on the marker in the map view, so I zoomed in and found many events that occurred in Israel. (...) It was interesting to find an event on the Suez Canal which I’d heard about but did not know the history of the Canal during World War I.''}

The event view is also effective in supporting event exploration and helping the participants find five events.
The main strategy for finding events with the event view was that they first chose some event charts based on interesting topic keywords, then investigated the important events (i.e., the pink dots) in each topic. 
P5 described how she found the desired event with this strategy: \textit{``I was reviewing the topic charts with the keyword `war.’ (...) I checked the dots and found an event that captured my attention (‘The Manhattan Project’).''} 
P6 also reported that \textit{``I saw topic explanations and found a war in Topic 1. (...) As the title of the event (`Glasgow ice cream wars’) seemed different from other events, I decided to read the article on the event out of curiosity. As I explored Topic 17 (Battle), I saw an event whose keyword `claymore’ caught my attention for reading—`Operation Claymore.'}'' 
The keyword search was used for exploring four events.
\\
\textbf{HisVA Supports the Generation of Questions on History:}
We found that as the participants started exploring, they quickly generated their own questions and hypotheses using HisVA. We saw that the participants reported producing 21 motivations using the event view (19 by topic keywords and 2 by summary chart) and 17 motivations with the map view. The participants also reported 16 motivations from using combinations of the views (map + list views, 9; topic keyword + list view, 4; map + event views, 3). Next, we describe how each view helped the users’ event exploration.
Next, we describe how each view helped the users’ event exploration.

The event view, which provides event charts, topics with keywords, and a summary chart, played the most important role in event exploration. We found several reasons for the frequent use of the event charts. First, the main topic and topic keywords play a role in the initial exploration. 
The participants looked over the topics and hovered on some event charts, checking topic keywords and exploring important events. 
In particular, when students saw a wide dot (i.e., aggregated multiple important events), they were immediately interested and zoomed to see how many events were represented.
For example, P5 reported that she looked over the topics with a focus on wars and battles in the World War II time range, then found a dot that piqued her interest (``The Manhattan Project'' in the Soviet-related topic). P18 stated that \textit{``the dots are helpful, as I can see an overall trend of important events for the topics by observing the dots in different time ranges.''}

We noted that the map view plays a crucial role in exploring historical events. 
One common exploration pattern using the map view was that participants began their exploration by panning and zooming the map to find places for further investigation. 
P13 commented on the use of the map during exploration: \textit{``The map view provided me with geographical information of locations where a series of events occurred. This (geographical) information helped me intuitively understand the relationships of the countries.''} 
The participants were often attracted by markers whose event numbers were very large or small. As P17 expressed, \textit{``I can click and easily investigate (with the markers) what I am interested in on the map. (...) I saw many events in Europe, so I mainly explored that area.''} 
The opposite case also existed, as P4 stated: \textit{``It was a surprise to me when I found an event in Antarctica on the map. (...) I found it fun to study that the event (`Project COLDFEET') was actually a secret operation of the CIA in Antarctica in 1962 against the Soviet Union.''}
\\
\textbf{HisVA Helps Event Exploration from Personal Experience:} 
Searching historical events without any context is difficult due to the lack of prior information.
We found curiosity from personal experience can prompt sufficient cues for finding an initial event for exploration. 
Once students understand the first event, they extend their investigation coverage to the regions adjacent to their current location on the map, rather than jumping to an event in another time range or region. 
We found 20 comments that describe various reasons generated from personal experience for exploration with HisVA. 
Examples include topics discussed with friends and places traveled. 
For example, P9 started her exploration from Taiwan because she remembered a question that she had in her travels in Taiwan. 
Thus, she first selected Taiwan on the map view, read the recommended articles that explained the relationships between Taiwan and neighboring countries (e.g., ``Tapani incident''), and studied  some contemporary history of Taiwan. 
She commented on her exploration: \textit{``After traveling in Taiwan twice, I have wondered why people in Taiwan have a preference for Japan, although Japan colonized Taiwan during World War II.''}
P12 explored events based on a story that her friend told. 
During the study, as she remembered her target event was related to Egypt and Israel, she extensively used the map and list views to search the event that occurred in Egypt, which is the ``Battle of Abu Tellul.'' 
P12 stated, \textit{``I like HisVA as I find the event my friend told me. I actually did not have information about the year or the name of the war, but HisVA helped me find it.''}

\subsection{User Score Analysis}

\begin{table}[t]
\begin{tabular} {m{0.3cm}  m{6cm} | m{1.2cm}}
    \# & {Question} & {mean score/7.0}  \\ \midrule
    Q1  & {HisVA helps me search for necessary resources} &  {5.5}   \\
    Q2  & {The map view is useful for the tasks}  & {6.37}  \\
    Q3  & {The summary chart is useful for the tasks}  & {5.34}\\ 
    Q4  & {The topic view is useful for the tasks}  & {4.5} \\
    Q5  & {HisVA is useful in studying  history}  & {5.5}    \\
    \bottomrule
\end{tabular}
\caption{\label{table_1}The post-questionnaire questions and scores. }
\vspace{-0.5cm}
\end{table}

In the post-experiment questionnaire, we asked five questions with reasons about the user’s experience with HisVA using a 7-point Likert scale (1, strongly negative to 7, strongly positive). 
Table~\ref{table_1} shows the questions and average scores. We asked Q1 to find out if HisVA helped the users feel that they could effectively search for learning materials (score: 5.5). Q2--Q4 were asked to evaluate how users evaluate each function of HisVA for the tasks \cleanRevision(scores: 6.37, 5.34, and 4.5 respectively). Q5 was to ascertain the overall effectiveness of HisVA (score: 5.5).

Overall, results show that HisVA provides necessary information to these participants for exploring historical events (Q1). 
Specifically, the map view most effectively supports users in navigating through what they needed to study history (Q2). 
There are two possible reasons for the high usefulness score for the map view. 
Maps are a visualization system that users feel familiar with and therefore find easy to use to access knowledge based on locations~\cite{Lee16}, which led to their frequent use of the map view. 
P10, who gave it 7 points, stated that \textit{``(using the map view) I could explore any country I wanted and intuitively choose desired events for further investigation.''}
Second, the map view provides overviews on events by regions and such overviews play an important role in guiding user exploration to gather information for the next stages of an investigation. P23 stated, \textit{``The map view allows me to have an overview of a series of events that occurred in the region, and I could estimate the relationships among the events based on the locations of the events.''} 
P23 also gave 7 points to the map view.

The event view received 4.92 points on average (summary chart, 5.34; topic charts, 4.5), which is higher than the neutral score (3.5).
We attribute this to the main role of the view--to reduce the event search space (R1). 
The summary chart supported users by providing the number of events by time in all the data, which points out which period is important in history. It also helps users efficiently narrow down the event search space with the vertical bars for filtering items by time in all other views. P18, who rated 7 points, stated: \textit{``I often became confused with events that occurred at the same time but in different regions. (...) In the event view, I could explore a series of events that occurred in a similar period for investigation.''} 
Note that he set the time range at 1941--1943 during his exploration and reported \textit{``Battle of Rzhev, Summer 1942,'' and “Battle of Midway,''} both of which happened in the same period but at different locations. The topic charts helped users find the themes in the given data and important entities (i.e., events in pink dots) that could guide exploration with the data. 
\textit{``Although I do not have knowledge of historical events, I could find events to focus on (with the dots) and the overall trends of events by both time and topics,''} as stated by P13, who gave 7 points to the event view. We also noticed that topic charts have a relatively lower score than the others (4.5). 
From our qualitative analysis, we surmise that some topics seemed similar to each other, so users may be confused in deciding which topics to focus on for their exploration. 
As we limited the number of topics to 20 in HisVA, some may not find what they expected, based on their personal knowledge or experience. 
P1, who gave 3 points to the event view, mentioned, \textit{``(...) It seems some topics look similar to each other, and the number of topics is less than I expected.''} 
We attribute this to the algorithm that models topics without supervision so we suspect that this issue can be alleviated by using models that allow user interventions for guided modeling~\cite{El-Assady20}.

Finally, we found that the participants thought that HisVA is an effective tool for studying history (Q5). Most participants mentioned that a strength of HisVA is the well-organized overviews provided that allow users to engage in efficient event exploration in both temporal and spatial contexts and an intuitive understanding of history.
As the tasks were designed by course instructors (E1 and E3 in Section~\ref{task_analysis}), this result is encouraging and shows HisVA's great potential to assist those who need to explore historical events in history courses.

\section{In-Class Deployment}
As we confirmed the potential of HisVA for training students to create stories by inferring relationships of events, we generated, distributed, and graded a take-home assignment in the history class (``History of Contemporary World'').
There were 25 undergraduate students enrolled in the class; they were mainly sophomores or juniors from diverse backgrounds (e.g., management, computer science, mechanical engineering, mathematics, physics). The students were assigned to one of five groups (five students/group) to complete the assignment. When the assignment was released, the students learned from lectures on imperialism, World War I and II, and the Cold War.

The assignment required each group to investigate relationships between the events during World Wars I and II (1910–1945) and the Cold War (1946–1990) with HisVA. 
When the assignment was distributed, the leading author of this work gave an in-class presentation with a (distributed) system manual and live demos of HisVA and answered students' questions. 
E3 also gave an instruction with examples that have not been discussed in the course (e.g., military conflicts in Africa and Southeast Asia). 
Note that the assignment instruction clearly stated that how to use the system was not part of the grading criteria, and the instructors did not consider the system use during the grading of the assignment. 
Instead, the instructors graded the submitted reports, focusing on the quality (e.g., how they creatively constructed their narratives) and findings acquired from their group activity.
As this was a take-home assignment for group work, the students were allowed to use any resources. They were guided to first prepare the assignment independently with the HisVA tool and then discuss individual work together in the class before the final submission, as they had done in their previous group assignments. 

Next, we report the three-fold positive outcomes and observations of using HisVA for studying history: 1) high-quality results, 2) diverse findings, and 3) self-directed learning.

\textbf{High quality results: }
The instructors reported that the students submitted different conflicts in many countries and understood the continuity of history from the U.S. and Soviet perspectives. 
Specifically, the students reported how colonized countries have survived and built independent nations, with many conflicts among domestic political parties after the end of colonialism. 
There were other insightful reports from students, such as how the newly built nations participated in the war between the United States and the Soviet Union (i.e., proxy war) due to a legacy of imperialism and colonialism. 
The instructors attributed this result to the functions of HisVA that allow both a ``global bird view'' and ``regionally-focused view.''
For example, a group of students remarked that they had investigated not only forests but also the trees inside the forests with HisVA. 
Another group reported that they had seen the continuity of events across various periods.

Reviewing the reports, we found that the map view played an important role in allowing not only detailed investigations of events by countries, but also connections between the investigation results and lectures' main points. For example, Group1 explored the events by countries in Southern Asia and then classified the events by four types of conflicts, which they studied during a lecture (conflicts between (1) imperial powers and Southeast Asian countries, (2) imperial powers, (3) the countries in the proxy wars, and (4) the countries in the Civil war). The group also reported that Thailand was never colonized in that era but indicated that the country experienced many coups, successfully presenting its own perspective on Thailand: ``(...) Thailand was a country at the time with an excellent ability to understand international political situations, but could not pay attention to internal problems, which led to coups and unstable domestic political conditions.''
The instructors stated that the result is encouraging, as it shows HisVA can be used to help students find new facts, relate the facts, and build their own perspectives of history.

\textbf{Diverse findings: } 
Students reported diverse events from many countries (e.g., Indonesia, Vietnam), which had not been discussed during lectures. 
They were irrelevant to the mainstream countries but had meanings in a regional context. 
For example, Group3 investigated ``Thai Nguyen uprising'' and ``Cochinchina uprising,'' which were not mainstream events but were important as they show South Vietnam’s history of resistance against the imperialism of France and Japan. 
The instructors confirmed that HisVA can provide students with opportunities to study history with a balanced perspective, stating: ``They (i.e., the reported diverse events) are rarely focused in history courses, but this (unpopularity) does not mean they are not important. (...) HisVA provides students with a chance to study other parts of the world that are also meaningful and connected to mainstream history.''

\textbf{Self-directed learning: } 
The instructors remarked that HisVA helps students think and connect different events to the main topic. 
We found that the discoveries and connections of events were the result of each team’s own choice of target regions, which was often made in the early exploration stage with multiple candidates. When deciding on target regions, students not only utilized knowledge acquired from lectures and personal experience but also actively used the information presented by HisVA. For example, Group2 chose Southeast Asia to investigate after discussion because they had studied about the Vietnam War in the history course, one of the crucial events in Southeast Asia during the Cold War, and the region is physically close to where the students lived. The students in this group also paid attention to the Middle East as well, although an unfamiliar region, as HisVA showed many events during the selected time range. Once they had initial target continents, they commenced using the main views in HisVA for further detailed investigation by country. This is similar to what we observed in the user study result—P1 and P6 voluntarily explored and interpreted the events in the regions of their choice, where only a few events existed in the event view.
According to the E2, these results indicate self-directed learning of history~\cite{Knowles75, Hammond14}, where students find their own questions and answers.

To sum up, the instructors concluded that HisVA has great potential and many strengths in comparing conventional approaches in supporting self-directed learning of history and complement existing materials that represent historians' pre-complied perspectives. 
One of the instructors stated: \textit{``What the students achieve is not simple acquisition of historical facts and cannot be performed with existing materials. (...) The students show the potential of performing critical and creative thinking in studying history by interactively and actively selecting, arranging, and interpreting events. (...) This result is encouraging as I have not seen this types of assignment reports in the past lectures without HisVA.''}

\section{Limitations and Discussion}
\label{sec_discussion}
In this section, we discuss limitations of the system in terms of scalability, topics, resources, and evaluation.

\textbf{Data Size and Addition:}
We used 3,019 Wikipedia articles that included time and location information. 
Although this number of articles included in the study might have been small, we think that it is sufficient for studying about historical events related to the World Wars, as shown in the usage scenarios and in-class deployment.
However, there could be a case wherein a larger number of articles is needed, such as studying about ancient history. Even with an increased data set for the case, we believe that HisVA can effectively support users with its interactive interfaces based on topics, computed importance, and aggregation approaches, although a long pre-processing time should be anticipated with increased topic counts. We think the best way to provide HisVA to users is to employ a set of focused topics and articles associated with course goals and topics; otherwise, students may become lost during their exploration.

\textbf{Topic Number and Transparency: }
Since a small fixed set of topics may limit students' exploration, it is possible to increase the number of topics based on a target coverage of history. Although HisVA supports a large number of topics and resource articles, due to its data-driven topic modeling approach and importance-based visual interfaces, we should note that incorporating a large number of topics may exacerbate the disorientation issue. Due to the black-box nature of the topic modeling method we used~\cite{Guidotti18}, there could be topics that are difficult for users to understand. To help users better understand such vague topics, the system could allow users to re-adjust the semantic relationships of the concepts~\cite{El-Assady20}, so that the users can summarize the historical collection and determine their next actions for future exploration. This approach could also complement the recommendations, allowing user-driven event exploration. In the future, we plan to apply the concept-based approach~\cite{park2017conceptvector} to allow students to build a historical concept of a specific object, phenomenon, or theme and then use it to explore and analyze the events. We also think that, if more users’ input was recorded, the system could improve the recommendations based on these inputs~\cite{Ricci15} and then assist in the exploration and interaction process, which would facilitate a more engaging learning experience.

\textbf{Using Other Resources To Complement Wikipedia}:  We used Wikipedia as a source for studying history, due to its openness and reliability, but the reliability of Wikipedia articles remains controversial, as described in Section~\ref{sec_rw_wiki}. As HisVA uses a data-driven approach in building a database for users, it is possible to add any desired resources to the system, such as textbooks and scholarly articles that complement Wikipedia articles~\cite{Rector08, Samoilenko17, Samoilenko18}.  In addition, adding new articles incrementally based on users’ requests could be considered~\cite{NIPS2010_3902} without the fear of overwhelming students.

\textbf{Evaluation}: 
We evaluated the usefulness of HisVA using two evaluation approaches: a user study and in-class deployment.
Although we show that HisVA is highly effective for students to study history, it is mainly based on qualitative analysis without a statistical test.
We plan to compare HisVA to web search (e.g., Google search) for quantitative evaluation (e.g., throughput, completion time, etc.) and to investigate statistical significance.

In this work, our aim is to design an effective history studying system; therefore, it is beyond the scope of this work to examine the effectiveness of the designed system compared to traditional pedagogical methods. As performing a formal comparison requires thorough investigation on many factors (e.g., lecture type, topic types) for experiment design, we leave this as future work.
As our target users are those who want to study history, we recruited the participants of the user study and the in-class deployment for an assignment from a history course. Thus, we could not completely rule out the possibility that the participants might have been reluctant to give negative feedback on the system, presuming adverse effects. However, considering that participants were notified that their feedback would not affect their final grades in any circumstances, such presumption is unlikely.
\section{Conclusion and Future Work}
We present HisVA to help users study history, collaborating with domain experts in history education. HisVA provides three visualization views—event, map, and resource, each of which helps users have an overview of historical events, reduce the exploration space, and find critical events. For evaluation, we conduct user studies, whose results indicate that HisVA effectively supports users' event exploration. 
We plan to design a visual interface for managing and sharing annotations and employing other data (e.g., an encyclopedia) for the full deployment of HisVA and an in-depth analysis of the event exploration processes (e.g., how users come up with different conclusions for a question).

\ifCLASSOPTIONcompsoc
  \section*{Acknowledgments}
  This work was supported by the National Research Foundation (NRF) grant (No. 2021R1A2C1004542, No. 2020R1H1A110101311). This work was also supported by Institute of Information \& communications Technology Planning \& Evaluation (IITP) grant funded by the Korea government (MSIT)-No.20200013360011001, Artificial Intelligence graduate school support (UNIST). The development and evaluation of the system and initial draft of this manuscript were done when Dongyun Han worked toward his M.S. degree at UNIST, Korea. 
  
\else
   regular IEEE prefers the singular form
  \section*{Acknowledgment}
\fi


\ifCLASSOPTIONcaptionsoff
  \newpage
\fi

\bibliographystyle{IEEEtran}
\bibliography{./bibtex/bib/0_reference}

\begin{thebibliography}{10}
\providecommand{\url}[1]{#1}
\csname url@samestyle\endcsname
\providecommand{\newblock}{\relax}
\providecommand{\bibinfo}[2]{#2}
\providecommand{\BIBentrySTDinterwordspacing}{\spaceskip=0pt\relax}
\providecommand{\BIBentryALTinterwordstretchfactor}{4}
\providecommand{\BIBentryALTinterwordspacing}{\spaceskip=\fontdimen2\font plus
\BIBentryALTinterwordstretchfactor\fontdimen3\font minus
  \fontdimen4\font\relax}
\providecommand{\BIBforeignlanguage}[2]{{%
\expandafter\ifx\csname l@#1\endcsname\relax
\typeout{** WARNING: IEEEtran.bst: No hyphenation pattern has been}%
\typeout{** loaded for the language `#1'. Using the pattern for}%
\typeout{** the default language instead.}%
\else
\language=\csname l@#1\endcsname
\fi
#2}}
\providecommand{\BIBdecl}{\relax}
\BIBdecl

\bibitem{levesque2015history}
S.~L{\'e}vesque and P.~Zanazanian, ````history is a verb: We learn it best when
  we are doing it!'': French and english canadian prospective teachers and
  history,'' \emph{Revista de Estudios Sociales}, no.~52, pp. 32--51, 2015.

\bibitem{mccarthy2000active}
J.~P. McCarthy and L.~Anderson, ``Active learning techniques versus traditional
  teaching styles: Two experiments from history and political science,''
  \emph{Innovative higher education}, vol.~24, no.~4, pp. 279--294, 2000.

\bibitem{brooks1999search}
J.~G. Brooks and M.~G. Brooks, \emph{In Search of Understanding: The Case for
  Constructivist Classrooms 2nd Edition}.\hskip 1em plus 0.5em minus
  0.4em\relax Pearson College Div, 2000.

\bibitem{clark2017surprise}
J.~Clark and A.~Nye, ``‘surprise me!’the (im) possibilities of agency and
  creativity within the standards framework of history education,''
  \emph{Educational Philosophy and Theory}, vol.~49, no.~6, pp. 656--668, 2017.

\bibitem{Seefeldt09}
D.~Seefeldt and W.~G. Thomas~III, ``What is digital history? a look at some
  exemplar projects,'' 2009.

\bibitem{richardson2003constructivist}
V.~Richardson, ``Constructivist pedagogy,'' \emph{Teachers college record},
  vol. 105, no.~9, pp. 1623--1640, 2003.

\bibitem{lebow1993constructivist}
D.~Lebow, ``Constructivist values for instructional systems design: Five
  principles toward a new mindset,'' \emph{Educational technology research and
  development}, vol.~41, no.~3, pp. 4--16, 1993.

\bibitem{driscoll1994psychology}
M.~P. Driscoll, \emph{Psychology of learning for instruction.}\hskip 1em plus
  0.5em minus 0.4em\relax Allyn \& Bacon, 1994.

\bibitem{krahenbuhl2016student}
K.~S. Krahenbuhl, ``Student-centered education and constructivism: Challenges,
  concerns, and clarity for teachers,'' \emph{The Clearing House: A Journal of
  Educational Strategies, Issues and Ideas}, vol.~89, no.~3, pp. 97--105, 2016.

\bibitem{poitras2014developing}
E.~G. Poitras and S.~P. Lajoie, ``Developing an agent-based adaptive system for
  scaffolding self-regulated inquiry learning in history education,''
  \emph{Educational Technology Research and Development}, vol.~62, no.~3, pp.
  335--366, 2014.

\bibitem{roberts2007state}
J.~C. Roberts, ``State of the art: Coordinated \& multiple views in exploratory
  visualization,'' in \emph{Fifth International Conference on Coordinated and
  Multiple Views in Exploratory Visualization (CMV 2007)}.\hskip 1em plus 0.5em
  minus 0.4em\relax IEEE, 2007, pp. 61--71.

\bibitem{savich2008improving}
C.~Savich, ``Improving critical thinking skills in history.'' \emph{Online
  Submission}, 2008.

\bibitem{cabiness2013integrating}
C.~Cabiness, L.~Donovan, and T.~D. Green, ``Integrating wikis in the support
  and practice of historical analysis skills,'' \emph{TechTrends}, vol.~57,
  no.~6, pp. 38--48, 2013.

\bibitem{boadu2014examination}
G.~Boadu, ``An examination of the use of technology in the teaching of history.
  a study of selected senior high schools in the cape coast metropolis,
  ghana.'' \emph{International Journal of Learning, Teaching and Educational
  Research}, vol.~8, no.~1, 2014.

\bibitem{Itoh12}
M.~Itoh and M.~Akaishi, ``Visualization for changes in relationships between
  historical figures in chronicles,'' in \emph{International Conference on
  Information Visualisation}, 2012, pp. 283--290.

\bibitem{Cho16}
I.~Cho, W.~Dou, D.~X. Wang, E.~Sauda, and W.~Ribarsky, ``Vairoma: A visual
  analytics system for making sense of places, times, and events in roman
  history,'' \emph{IEEE transactions on visualization and computer graphics},
  vol.~22, no.~1, pp. 210--219, 2016.

\bibitem{Firat18}
E.~E. F{\i}rat and R.~S. Laramee, ``{Towards a Survey of Interactive
  Visualization for Education},'' in \emph{Proceedings of Computer Graphics and
  Visual Computing (CGVC)}.\hskip 1em plus 0.5em minus 0.4em\relax The
  Eurographics Association, 2018, pp. 91--101.

\bibitem{Rector08}
L.~H. Rector, ``Comparison of wikipedia and other encyclopedias for accuracy,
  breadth, and depth in historical articles,'' \emph{Reference services
  review}, vol.~36, pp. 7--22, 2008.

\bibitem{Samoilenko17}
A.~Samoilenko, F.~Lemmerich, K.~Weller, M.~Zens, and M.~Strohmaier, ``Analysing
  timelines of national histories across wikipedia editions: A comparative
  computational approach,'' in \emph{Eleventh International AAAI Conference on
  Web and Social Media}, 2017, pp. 210--219.

\bibitem{Samoilenko18}
A.~Samoilenko, F.~Lemmerich, M.~Zens, M.~Jadidi, M.~G{\'e}nois, and
  M.~Strohmaier, ``(don't) mention the war: A comparison of wikipedia and
  britannica articles on national histories,'' in \emph{International World
  Wide Web Conferences}, 2018, pp. 843--852.

\bibitem{Dimitrov17}
D.~Dimitrov, P.~Singer, F.~Lemmerich, and M.~Strohmaier, ``What makes a link
  successful on wikipedia?'' in \emph{International Conference on World Wide
  Web}.\hskip 1em plus 0.5em minus 0.4em\relax International World Wide Web
  Conferences, 2017, pp. 917--926.

\bibitem{Schwarzer16}
M.~Schwarzer, M.~Schubotz, N.~Meuschke, C.~Breitinger, V.~Markl, and B.~Gipp,
  ``Evaluating link-based recommendations for wikipedia,'' in \emph{2016
  IEEE/ACM Joint Conference on Digital Libraries (JCDL)}.\hskip 1em plus 0.5em
  minus 0.4em\relax IEEE, 2016, pp. 191--200.

\bibitem{Nguyen18}
P.~H. Nguyen, C.~Turkay, G.~Andrienko, N.~Andrienko, O.~Thonnard, and
  J.~Zouaoui, ``Understanding user behaviour through action sequences: from the
  usual to the unusual,'' \emph{IEEE transactions on visualization and computer
  graphics}, vol.~25, no.~9, pp. 2838--2852, 2018.

\bibitem{Ko12}
S.~Ko, R.~Maciejewski, Y.~Jang, and D.~S. Ebert, ``Marketanalyzer: An
  interactive visual analytics system for analyzing competitive advantage using
  point of sale data,'' vol.~31, no.~3, 2012, pp. 1245--1254.

\bibitem{Liu17b}
Z.~Liu, Y.~Wang, M.~Dontcheva, M.~Hoffman, S.~Walker, and A.~Wilson, ``Patterns
  and sequences: Interactive exploration of clickstreams to understand common
  visitor paths,'' \emph{IEEE Transactions on Visualization and Computer
  Graphics}, vol.~23, no.~1, pp. 321--330, 2017.

\bibitem{Monroe13}
M.~Monroe, R.~Lan, H.~Lee, C.~Plaisant, and B.~Shneiderman, ``Temporal event
  sequence simplification,'' \emph{IEEE transactions on visualization and
  computer graphics}, vol.~19, no.~12, pp. 2227--2236, 2013.

\bibitem{Dou12}
W.~Dou, X.~Wang, D.~Skau, W.~Ribarsky, and M.~X. Zhou, ``Leadline: Interactive
  visual analysis of text data through event identification and exploration,''
  in \emph{IEEE Conference on Visual Analytics Science and Technology}.\hskip
  1em plus 0.5em minus 0.4em\relax IEEE, 2012, pp. 93--102.

\bibitem{Shi15}
C.~Shi, S.~Fu, Q.~Chen, and H.~Qu, ``Vismooc: Visualizing video clickstream
  data from massive open online courses,'' in \emph{IEEE Pacific visualization
  symposium}, 2015, pp. 159--166.

\bibitem{chen2018viseq}
Q.~Chen, X.~Yue, X.~Plantaz, Y.~Chen, C.~Shi, T.-C. Pong, and H.~Qu, ``Viseq:
  Visual analytics of learning sequence in massive open online courses,''
  \emph{IEEE transactions on visualization and computer graphics}, 2018.

\bibitem{Bergmann12}
J.~Bergmann and A.~Sams, \emph{Flip your classroom: Reach every student in
  every class every day}.\hskip 1em plus 0.5em minus 0.4em\relax International
  society for technology in education, 2012.

\bibitem{O15}
J.~O'Flaherty and C.~Phillips, ``The use of flipped classrooms in higher
  education: A scoping review,'' \emph{The internet and higher education},
  vol.~25, pp. 85--95, 2015.

\bibitem{Kim14}
M.~K. Kim, S.~M. Kim, O.~Khera, and J.~Getman, ``The experience of three
  flipped classrooms in an urban university: an exploration of design
  principles,'' \emph{The Internet and Higher Education}, vol.~22, pp. 37--50,
  2014.

\bibitem{William01}
J.~WILLIAM P.~Eveland and S.~Dunwoody, ``User control and structural
  isomorphism or disorientation and cognitive load?: Learning from the web
  versus print,'' \emph{Communication research}, vol.~28, no.~1, pp. 48--78,
  2001.

\bibitem{Paulo99}
P.~Dias, M.~J. Gomes, and A.~P. Correia, ``Disorientation in hypermedia
  environments: Mechanisms to support navigation,'' \emph{Journal of
  Educational Computing Research}, vol.~20, no.~2, pp. 93--117, 1999.

\bibitem{Westad05}
O.~A. Westad, \emph{The Global Cold War}.\hskip 1em plus 0.5em minus
  0.4em\relax Cambridge University Press, 2005.

\bibitem{andrienko2010space}
G.~Andrienko, N.~Andrienko, S.~Bremm, T.~Schreck, T.~Von~Landesberger, P.~Bak,
  and D.~Keim, ``Space-in-time and time-in-space self-organizing maps for
  exploring spatiotemporal patterns,'' in \emph{Computer Graphics Forum},
  vol.~29, no.~3.\hskip 1em plus 0.5em minus 0.4em\relax Wiley Online Library,
  2010, pp. 913--922.

\bibitem{Bizer09}
C.~Bizer, J.~Lehmann, G.~Kobilarov, S.~Auer, C.~Becker, R.~Cyganiak, and
  S.~Hellmann, ``Dbpedia - a crystallization point for the web of data,''
  \emph{Journal of Web Semantics}, vol.~7, no.~3, pp. 154 -- 165, 2009.

\bibitem{Wiki_WW1}
\BIBentryALTinterwordspacing
``World war i,'' accessed: 2019-08-25. [Online]. Available:
  \url{https://en.wikipedia.org/wiki/World_War_I}
\BIBentrySTDinterwordspacing

\bibitem{Wiki_WW2}
\BIBentryALTinterwordspacing
``World war ii,'' accessed: 2019-08-25. [Online]. Available:
  \url{https://en.wikipedia.org/wiki/World_War_II}
\BIBentrySTDinterwordspacing

\bibitem{Wiki_COLDWAR}
\BIBentryALTinterwordspacing
``Cold war,'' \url{https://en.wikipedia.org/wiki/Cold_war}, accessed:
  2019-08-25. [Online]. Available: \url{https://en.wikipedia.org/wiki/Cold_war}
\BIBentrySTDinterwordspacing

\bibitem{Finkel05}
J.~R. Finkel, T.~Grenager, and C.~Manning, ``Incorporating non-local
  information into information extraction systems by gibbs sampling,'' in
  \emph{Association for Computational Linguistics}, 2005, pp. 363--370.

\bibitem{grishman1996message}
R.~Grishman and B.~M. Sundheim, ``Message understanding conference-6: A brief
  history,'' in \emph{COLING 1996 Volume 1: The 16th International Conference
  on Computational Linguistics}, 1996.

\bibitem{Geopy06}
\BIBentryALTinterwordspacing
``Geopy,'' \url{https://readthedocs.org/projects/geopy/}, 2006. [Online].
  Available: \url{https://readthedocs.org/projects/geopy/}
\BIBentrySTDinterwordspacing

\bibitem{hofmann1999probabilistic}
T.~Hofmann, ``Probabilistic latent semantic indexing,'' in \emph{Proceedings of
  the 22nd annual international ACM SIGIR conference on Research and
  development in information retrieval}, 1999, pp. 50--57.

\bibitem{teh2005sharing}
Y.~W. Teh, M.~I. Jordan, M.~J. Beal, and D.~M. Blei, ``Sharing clusters among
  related groups: Hierarchical dirichlet processes,'' in \emph{Advances in
  neural information processing systems}, 2005, pp. 1385--1392.

\bibitem{blei2003latent}
D.~M. Blei, A.~Y. Ng, and M.~I. Jordan, ``Latent dirichlet allocation,''
  \emph{Journal of machine Learning research}, vol.~3, no. Jan, pp. 993--1022,
  2003.

\bibitem{McCallum02}
A.~K. McCallum, ``Mallet: A machine learning for language toolkit,'' 2002,
  http://mallet.cs.umass.edu.

\bibitem{roder2015exploring}
M.~R{\"o}der, A.~Both, and A.~Hinneburg, ``Exploring the space of topic
  coherence measures,'' in \emph{ACM international conference on Web search and
  data mining}, 2015, pp. 399--408.

\bibitem{rehurek_lrec}
R.~{\v R}eh{\r u}{\v r}ek and P.~Sojka, ``{Software Framework for Topic
  Modelling with Large Corpora},'' in \emph{{LRECWorkshop on New Challenges for
  NLP Frameworks}}.\hskip 1em plus 0.5em minus 0.4em\relax ELRA, 2010, pp.
  45--50.

\bibitem{Whitelaw15}
W.~Mitchell, ``Generous interfaces for digital cultural collections,''
  \emph{Digital Humanities Quarterly}, vol.~9, no.~1, 2015.

\bibitem{Mayr16}
E.~Mayr, P.~Federico, S.~Miksch, G.~Schreder, M.~Smuc, and F.~Windhager,
  ``Visualization of cultural heritage data for casual users,'' in \emph{IEEE
  VIS Workshop on Visualization for the Digital Humanities}, 2016.

\bibitem{Windhager18}
F.~Windhager, P.~Federico, G.~Schreder, K.~Glinka, M.~D{\"o}rk, S.~Miksch, and
  E.~Mayr, ``Visualization of cultural heritage collection data: State of the
  art and future challenges,'' \emph{IEEE transactions on visualization and
  computer graphics}, vol.~25, pp. 2311--2330, 2018.

\bibitem{D3js}
\BIBentryALTinterwordspacing
``D3js,'' accessed: 2019-08-25. [Online]. Available: \url{http://www.d3js.org}
\BIBentrySTDinterwordspacing

\bibitem{Leafletjs}
\BIBentryALTinterwordspacing
``Leaflet,'' accessed: 2019-08-25. [Online]. Available:
  \url{https://leafletjs.com/}
\BIBentrySTDinterwordspacing

\bibitem{Flask10}
\BIBentryALTinterwordspacing
``Flask,'' accessed: 2019-09-01. [Online]. Available:
  \url{https://flask.palletsprojects.com/en/1.0.x/}
\BIBentrySTDinterwordspacing

\bibitem{Leaflet_MarkerCluster}
\BIBentryALTinterwordspacing
``Leaflet markercluster,'' accessed: 2019-08-25. [Online]. Available:
  \url{https://github.com/Leaflet/Leaflet.markercluster}
\BIBentrySTDinterwordspacing

\bibitem{ware2019information}
C.~Ware, \emph{Information visualization: perception for design}.\hskip 1em
  plus 0.5em minus 0.4em\relax Morgan Kaufmann, 2019.

\bibitem{bertin1983semiology}
J.~Bertin, ``Semiology of graphics; diagrams networks maps,'' Tech. Rep., 1983.

\bibitem{lee2019visual}
C.~Lee, Y.~Kim, S.~M. Jin, D.~Kim, R.~Maciejewski, D.~Ebert, and S.~Ko, ``A
  visual analytics system for exploring, monitoring, and forecasting road
  traffic congestion,'' \emph{IEEE Transactions on Visualization and Computer
  Graphics}, vol.~26, no.~11, pp. 3133--3146, 2020.

\bibitem{Liang09}
H.-N. Liang and K.~Sedig, ``Characterizing navigation in interactive learning
  environments,'' \emph{Interactive Learning Environments}, vol.~17, no.~1, pp.
  53--75, 2009.

\bibitem{Alpha_History}
\BIBentryALTinterwordspacing
``alpha history,'' accessed: 2019-08-25. [Online]. Available:
  \url{https://alphahistory.com/worldwar1/world-war-i-essay-questions/}
\BIBentrySTDinterwordspacing

\bibitem{Hammond14}
M.~Hammond and R.~Collins, \emph{Self-directed Learning: Critical
  Practice}.\hskip 1em plus 0.5em minus 0.4em\relax Routledge Falmer Taylor and
  Francis Group, 2014.

\bibitem{thomas2003general}
D.~R. Thomas, ``A general inductive approach for qualitative data analysis,''
  2003.

\bibitem{Lee16}
S.~Lee, S.-H. Kim, and B.~C. Kwon, ``Vlat: Development of a visualization
  literacy assessment test,'' \emph{IEEE transactions on visualization and
  computer graphics}, vol.~23, no.~1, pp. 551--560, 2016.

\bibitem{El-Assady20}
M.~El-Assady, R.~Kehlbeck, C.~Collins, D.~Keim, and O.~Deussen, ``Semantic
  concept spaces: Guided topic model refinement using word-embedding
  projections,'' \emph{IEEE transactions on visualization and computer
  graphics}, vol.~26, no.~1, pp. 1001--1011, 2019.

\bibitem{Knowles75}
M.~Knowles, \emph{Self-Directed Learning: A Guide for Learners and
  Teachers}.\hskip 1em plus 0.5em minus 0.4em\relax Association Press, 1975.

\bibitem{Guidotti18}
R.~Guidotti, A.~Monreale, S.~Ruggieri, F.~Turini, F.~Giannotti, and
  D.~Pedreschi, ``A survey of methods for explaining black box models,''
  \emph{ACM Computing Survey}, vol.~51, no.~5, pp. 93:1--93:42, 2019.

\bibitem{park2017conceptvector}
D.~Park, S.~Kim, J.~Lee, J.~Choo, N.~Diakopoulos, and N.~Elmqvist,
  ``Conceptvector: text visual analytics via interactive lexicon building using
  word embedding,'' \emph{IEEE transactions on visualization and computer
  graphics}, vol.~24, no.~1, pp. 361--370, 2017.

\bibitem{Ricci15}
B.~S. Francesco~Ricci, Lior~Rokach, \emph{Recommender Systems Handbook}.\hskip
  1em plus 0.5em minus 0.4em\relax Springer, 2015.

\bibitem{NIPS2010_3902}
\BIBentryALTinterwordspacing
M.~Hoffman, F.~R. Bach, and D.~M. Blei, ``Online learning for latent dirichlet
  allocation,'' in \emph{Advances in Neural Information Processing Systems 23},
  J.~D. Lafferty, C.~K.~I. Williams, J.~Shawe-Taylor, R.~S. Zemel, and
  A.~Culotta, Eds.\hskip 1em plus 0.5em minus 0.4em\relax Curran Associates,
  Inc., 2010, pp. 856--864. [Online]. Available:
  \url{http://papers.nips.cc/paper/3902-online-learning-for-latent-dirichlet-allocation.pdf}
\BIBentrySTDinterwordspacing

\end{thebibliography}

\begin{IEEEbiographynophoto}
{Dongyun Han}
is a Ph.D. student in the department of computer science at Utah State University. His research interests include Visual Analytics and Virtual Environments. He received a M.S in Computer Science at UNIST (Ulsan National Institute of Science and Technology) in 2020.
\end{IEEEbiographynophoto}

\begin{IEEEbiographynophoto}
{Gorakh Parsad}
is a research graduate student (M.S) in department of Computer Science at UNIST, South Korea. His research interests include data visualization, A.I. applications domain and HCI.
\end{IEEEbiographynophoto}

\begin{IEEEbiographynophoto}
{Hwiyeon Kim}
is a research graduate student (M.S) in department of Computer Science at UNIST, South Korea. Her research interests include data journalism and HCI.
\end{IEEEbiographynophoto}

\begin{IEEEbiographynophoto}
{Jaekyom Shim} is working at the Leadership Center at UNIST. He worked for the project, ``Korea and East Asia in Global History, 1840-2000,'' from 2011 to 2014, at the Friedrich-Meinecke-Institut, Frei Universität Berlin, and was a visiting fellow at the Institute for Advanced Studies on Asia, University of Tokyo, as a beneficiary of Japanese Studies Overseas Fellowship of Japan Foundation. His main concern is history of postwar East Asia and transnational history of popular culture.
\end{IEEEbiographynophoto}

\begin{IEEEbiographynophoto}
{Oh-Sang Kwon}is an associate professor in the department of biomedical engineering at UNIST, South Korea. His research interests include human visual perception, perceptual/cognitive biases, perception-action interaction, and perceptual learning. He received the doctoral degree in psychological sciences at Purdue University in 2009. For more information, visit http://pal.unist.ac.kr
\end{IEEEbiographynophoto}

\begin{IEEEbiographynophoto}
{Kyung A Son} is a research professor in The U Innovation Education Center at UNIST, South Korea. Her research interests include educational technology, instructional design for leaner-oriented education, development of teaching and learning model and strategies, and planning educational innovation and policy. She received her Ph.D. in Educational Technology from Hanyang University in 2003.   
\end{IEEEbiographynophoto}

\begin{IEEEbiographynophoto}
{Jooyoung Lee} is an assistant professor in the School of Liberal Arts at UNIST, Korea. His fields of research include American diplomacy, U.S. in the world, history of human rights, history of colonialism, anti-colonialism and post-colonial nation-building, and cultural history. Having an interest in the innovation of  history education, he has been trying various ways to engage students and create a learner-centered environment. He received his Ph.D. in history from Brown University in 2012. 
\end{IEEEbiographynophoto}

\begin{IEEEbiographynophoto}
{Isaac Cho}
is an assistant professor in the computer science department at Utah State University and an adjunct professor at the Computer Science Department at the University of North Carolina at Charlotte. His main research interests are Interactive Visual Analytics, User Interfaces,  and Virtual Environments. He received a Ph.D. in Computing and Information Systems from the University of North Carolina at Charlotte in 2013.
\end{IEEEbiographynophoto}

\begin{IEEEbiographynophoto}
{Sungahn Ko} is an associate professor in the School of Computer Science and Engineering at UNIST, Ulsan, South Korea. His research interests include visual analytics, information visualization, and Human-Computer Interaction. He received the doctoral degree in electrical and computer engineering from Purdue University in 2014. For more information, visit http://ivader.unist.ac.kr
\end{IEEEbiographynophoto}





\end{document}